\documentclass[seceq]{ptptex}
\usepackage{wrapft}

\usepackage[dvips]{graphicx}% Include figure files
\usepackage{dcolumn}% Align table columns on decimal point
\usepackage{bm}% bold math

\def\mapgeq{\mathbin{\lower.3ex\hbox{$\buildrel>\over{\smash{\scriptstyle\sim}\vphantom{_x}}$}}}
\def\mapleq{\mathbin{\lower.3ex\hbox{$\buildrel<\over{\smash{\scriptstyle\sim}\vphantom{_x}}$}}}
\def\mapgeqeq{\mathbi{\lower.3ex\hbox{$\buildrel>\over{\smash{\scriptstyle\approx}\vphantom{_2}}$}}}
\def\mapleqeq{\mathbin{\lower.3ex\hbox{$\buildrel<\over{\smash{\scriptstyle\approx}\vphantom{_2}}$
}}}
\def\Journal#1#2#3#4{{#1} {\bf #2} (#4) #3}
\def\NPB{Nucl. Phys. B}

\def\NPSUPPL{Nucl. Phys. Proc. Suppl.}
\def\PLB{{Phys. Lett.} B}

\def\PLBOLD{Phys. Lett.}
\def\PRL{Phys. Rev. Lett.}
\def\RMP{Rev. Mod. Phys.}
\def\PRD{Phys. Rev. D}

\def\PTP{Prog. Theor. Phys.}
\def\JHEP{JHEP}

\def\EPJ{Euro. Phys. J. C}

\def\JETPUSSR{Sov. Phys. JETP}

\def\ZETP{Zh. Eksp. Teor. Fiz.}

\def\IJMPE{Int. J. Mod. Phys. E}

\def\SCI{Science}
\def\APJ{Astrophysics J.}

\def\NJP{New. J. Phys.}

\def\Erratum{Erratum-ibid}

\title{%        %You can use \\ for explicit line-break.
Majorana CP Violation in Approximately $\mu$-$\tau$ Symmetric Models with det(M$_\nu$)=0
}
\author{%       %Use \scshape  for the family name.
Teppei \textsc{Baba}\footnote{E-mail: 7atrd014@keyaki.cc.u-tokai.ac.jp} 
and Masaki \textsc{Yasu\`{e}}\footnote{E-mail: yasue@keyaki.cc.u-tokai.ac.jp}
}

\inst{%         %Affiliation, neglected when [addenda] or [errata].
Department of Physics, Tokai University,\\
1117 Kitakaname, Hiratsuka, Kanagawa 259-1292, Japan
}

\abst{%         %This abstract is neglected when [addenda] or [errata].
We discuss effects of Majorana CP violation in a model-independent way 
for a given phase structure of flavor neutrino masses. 
To be more predictive, we confine ourselves to models with $\det(M_\nu)=0$, where $M_\nu$ is a 
flavor neutrino mass matrix, and to be consistent with observed results of the neutrino oscillation,
the models are subject to an approximate $\mu$-$\tau$ symmetry.  There are two categories of 
approximately $\mu$-$\tau$ symmetric models classified as (C1) yielding 
$\sin^22\theta_{23} \approx 1$ and $\sin^2\theta_{13} \ll 1$ and (C2) 
yielding $\sin^22\theta_{23} \approx 1$ 
and $\Delta m_\odot^2/\vert\Delta m_{atm}^2\vert\ll 1$, where $\theta_{23(13)}$ stands for the 
mixing of massive neutrinos 
$\nu_2$ and $\nu_3$ ($\nu_1$ and $\nu_3$) and $\Delta m_ \odot ^2$ ($\Delta m_{atm}^2$) stands for 
the mass squared difference 
for atmospheric (solar) neutrinos. 
The Majorana phase can be large for the normal mass hierarchy and for
the inverted mass hierarchy with $m_1\approx -m_2$ only realized in (C1)
while they are generically small for the inverted mass hierarchy with $m_1\approx m_2$ in both (C1) and (C2).  
These results do not 
depend on a specific choice of phases in $M_\nu$ but hold true in any models with $\det(M_\nu)=0$ because of the 
rephasing invariance.
}
\begin{document}
\maketitle

%%%%% Abstract %%%%%%%%%%%%%%%%%%%
\section{\label{sec:1}Introduction}
Neutrinos are oscillating and mixed with each other among three flavor neutrinos.  Such oscillations have been 
confirmed to occur for the atmospheric neutrinos \cite{atmospheric-neutrinos}, 
the solar neutrinos \cite{old-solar-neutrinos,solar-neutrinos}, 
the reactor neutrinos \cite{reactor-neutrinos} and 
the accelerator neutrinos \cite{accelerator-neutrinos}.  Three massive neutrinos have masses 
$m_{1,2,3}$ measured as mass squared differences defined by 
$\Delta m^2_\odot = m^2_2-m^2_1$ and $\Delta m^2_{atm} = m^2_3-m^2_1$.  
Three flavor neutrinos $\nu_{e,\mu,\tau}$ are mixed into three massive neutrinos $\nu_{1,2,3}$ 
during their flight and the mixing can be 
described by the Pontecorvo-Maki-Nakagawa-Sakata (PMNS) mixing matrix \cite{PMNS} parameterized by 
three mixing angles $\theta_{12,23,13}$, one Dirac CP violating phase $\delta_{CP}$ and three Majorana phase 
$\phi_{1,2,3}$ \cite{CP-Violation}, where Majorana CP violating phases are given by two combinations of $\phi_{1,2,3}$.

It is CP property of neutrinos that has currently received much attention since the similar CP property 
of quarks has been observed and successfully described by the Kobayashi-Maskawa mixing matrix 
\cite{Kobayashi-Maskawa}. If neutrinos exhibit CP violation, there is a new seed 
to produce the baryon number in the Universe 
by the Fukugida-Yanagida mechanism of the leptogenesis \cite{leptogenesis}, which favors 
the seesaw mechanism \cite{seesaw-model} of creating tiny 
neutrino masses.  However, there is no direct linkage between 
CP violation of three flavor neutrinos and that of the leptogenesis since 
the CP violating phases are associated with heavy neutrinos but not with three flavor neutrinos.
If the number of the heavy neutrinos is two, there is one-to-one correspondence between the CP violating phases 
of flavor neutrinos and that of the seesaw mechanism.  The model with two heavy neutrinos is called 
minimal seesaw model \cite{minimal-seesaw-model}. 
Even if the minimal seesaw model is adopted, predictions of the leptonic CP violation depend on 
the choice of various parameters including Dirac neutrino mass terms.  It seems of great significance to make
predictions independent of the specific parameter choice.  The general feature of 
the minimal seesaw model is that it satisfies $\det(M_\nu)=0$, where $M_\nu$ represents a 
flavor neutrino mass matrix.  Therefore, we choose this condition $\det(M_\nu)=0$ as our standing 
point to investigate effects of leptonic CP violations as general as possible.

To discuss the leptonic CP violation starting from a given phase structure of $M_\nu$, 
we have to clarify how phases of flavor neutrino masses affect the leptonic CP phases.
To do so, we have to mathematically consider 6 phase parameters in the PMNS mixing matrix $U_{PMNS}$ 
to cover a general phase structure of $M_\nu$. It should be noted that 
conventional studies utilizing the standard version of 
the PMNS matrix $U^{PDG}_{PMNS}$ given by the Particle Data Group (PDG) \cite{PDG}
 and $m_{1,2,3}$ do not provide a clue to see direct effects from 
phases of the flavor neutrino masses \cite{Conventional}.  In our study, $U_{PMNS}$ is parameterized by 
$U_{PMNS}=UK$ with
%%%%%%%%%%%%%%%%%%%%
\begin{eqnarray}
 U &=& \left( {\begin{array}{*{20}c}
   1 & 0 & 0  \\
   0 & {e^{i\gamma } } & 0  \\
   0 & 0 & {e^{ - i\gamma } }  \\
\end{array}} \right)
\left( {\begin{array}{*{20}c}
   1 & 0 & 0  \\
   0 & {c_{23} } & {s_{23} }  \\
   0 & { - s_{23} } & {c_{23} }  \\
\end{array}} \right)
\left( {\begin{array}{*{20}c}
   {c_{13} } & 0 & {s_{13} e^{ - i\delta } }  \\
   0 & 1 & 0  \\
   { - s_{13} e^{i\delta } } & 0 & {c_{13} }  \\
\end{array}} \right)
\nonumber\\
&&\cdot\left( {\begin{array}{*{20}c}
   {c_{12} } & {s_{12} e^{i\rho } } & 0  \\
   { - s_{12} e^{ - i\rho } } & {c_{12} } & 0  \\
   0 & 0 & 1  \\
\end{array}} \right),
\label{Eq:PMNS matrix}\\
K &=& {\rm diag.}\left( {e^{i\varphi _1 },e^{i\varphi _2 },e^{i\varphi _3 } } \right), 
\nonumber
\end{eqnarray}
%%%%%%%%%%%%%%%%%%%%
where $c_{ij}=\cos\theta_{ij}$ and $s_{ij}=\sin\theta_{ij}$ ($i,j$=1,2,3). 
There is the similar phase to $\delta$ and $\rho$, say $\chi$, 
which can be associated with the $\nu_2$-$\nu_3$ mixing.  However, we have confirmed that the phase 
of the $\nu_2$-$\nu_3$ mixing should be $\gamma$ in place of $\chi$ to consistently describe 
the neutrino mixings \cite{Dependence-of-CP-phase-on-X-and-Y}.  Namely, Eq.(\ref{Eq:PMNS matrix})
 with 
$\chi$ 
included is converted into $UK$ with ${\tilde \delta}=\delta+\chi/2$, ${\tilde \rho}=\rho+\chi/2$, 
${\tilde \gamma}=\gamma+\chi/2$,
${\tilde \varphi_2}=\varphi_2+\chi/2$ and ${\tilde \varphi_3}=\varphi_3+\chi/2$ as obvious 
replacements.
Physically, among $\delta$, $\rho$, and $\gamma$ two phases are redundant. By defining 
$\phi_1=\varphi_1-\rho$ (as well as $\phi_{2,3}=\varphi_{2,3}$) and 
%%%%%%%%%%%%%%%%%%%%
\begin{equation}
\delta_{CP}=\delta + \rho, 
\label{Eq:delta_CP}
\end{equation}
%%%%%%%%%%%%%%%%%%%%
we reach $U^{PDG}_{PMNS}=U^{PDG}K^{PDG} $ consisting of
%%%%%%%%%%%%%%%%%%%%
\begin{eqnarray}
 U^{PDG} &=& \left( {\begin{array}{*{20}c}
   {c_{12} c_{13} } & {s_{12} c_{13} }  \\
   { - c_{23} s_{12}  - s_{23} c_{12} s_{13} e^{i\delta _{CP} } } & {c_{23} c_{12}  
   - s_{23} s_{12} s_{13} e^{i\delta _{CP} } }  \\
   {s_{23} s_{12}  - c_{23} c_{12} s_{13} e^{i\delta _{CP} } } & { - s_{23} c_{12}  
   - c_{23} s_{12} s_{13} e^{i\delta _{CP} } }  \\
\end{array}}
{\begin{array}{*{20}c}
   {s_{13} e^{ - i\delta _{CP} } }  \\
   {s_{23} c_{13} }  \\
   {c_{23} c_{13} }  \\
\end{array}} \right),
\nonumber\\
 K^{PDG} &=& {\rm diag.}\left( {e^{i\phi _1  },e^{i\phi _2 },e^{i\phi _3 } } \right).
\label{Eq:PDG version PMNS matrix}
\end{eqnarray}
%%%%%%%%%%%%%%%%%%%%
Furthermore, one Majorana phase is also redundant and the CP violating Majorana phases are given by 
two combinations of the Majorana phases such as $\phi_{i}-\phi_{3}$ ($i=1,2$). For the reader's 
convenience, we show three typical forms of $U_{PMNS}$ in Appendix \ref{sec:Appendix-rephasing}.

%%%%%%%%%%%%Introduction and Explanation of previous work%%%%%%%%%%%%
The neutrino mixing angles and mass squared differences have been measured by recent neutrino 
oscillation experiments.
The current data of the mixing angles and mass squared differences are shown as \cite{current-data}:
%%%%%%%%%%%%%%%%%%%%
\begin{eqnarray}
&&
\sin ^2 \theta _{12}  = 0.30_{ - 0.02}^{+ 0.02},
 \quad
 \sin ^2 \theta _{23}  =  0.50_{- 0.06}^{ + 0.07},
 \quad
 \sin ^2 \theta _{13}  < 0.040,  
\nonumber\\ 
&& 
\Delta m_ \odot ^2  = \left( 7.65_{ - 0.20}^{ + 0.23} \right) \times 10^{ - 5}~{\rm eV}^2,
 \quad
 \left| {\Delta m_{atm}^2 } \right| = \left(2.40_{-0.11}^{+0.12} \right) \times 10^{ - 3}~{\rm eV}^2.
\label{Eq:experimental data}
\end{eqnarray}
%%%%%%%%%%%%%%%%%%%%
The gross property of the experimental data indicating the almost maximal atmospheric neutrino mixing and 
the small 1-3 neutrino mixing can be understood as a result of the $\mu$-$\tau$ symmetry \cite{mu-tau} imposed 
on neutrino interactions, which gives $\sin^2\theta_{23}=0.5$ and $\sin^2\theta_{13}=0$.  However, there is 
no Dirac CP violation. If Dirac CP violation is observed in future neutrino experiments \cite{Dirac-CP-violation-is-observed}, 
we have to include 
tiny violation of the $\mu$-$\tau$ symmetry.\footnote{The $\mu$-$\tau$ symmetry breaking should be 
present because the charged leptons placed into $SU(2)_L$-doublets together with the flavor neutrinos 
violate the $\mu$-$\tau$ symmetry.}  
If the $\mu$-$\tau$ symmetry breaking is included, 
there are two categories of textures respectively referred to as (C1) and (C2) 
\cite{C1-C2}.
In the category (C1), we have $\sin 2\theta _{23}  \approx \sigma$ ($\sigma = \pm 1$) and $\sin^2\theta_{13}\ll1$
 while in the category (C2), we have $\sin 2\theta _{23}  \approx - \sigma$ and 
$\Delta m^2_ \odot/\vert\Delta m^2_{atm}\vert \ll 1$.
In the category (C2), the $\mu$-$\tau$ symmetric limit is signaled by $\sin\theta_{12}\rightarrow 0$ instead of 
$\sin\theta_{13}\rightarrow 0$.  The phenomenologically consistent value of $\sin\theta_{12}$ is 
realized by the form of $\tan 2\theta_{12}\propto \varepsilon/\eta$, where $\varepsilon$ represents the 
$\mu$-$\tau$ symmetry breaking parameter and another small parameter denoted by $\eta$ of 
${\mathcal O}(\varepsilon)$ is required.

%%%%%%%%%%%%%%%%%%%%%%%%%%%% ˜_•¶"à'Å'â'邱'Æ %%%%%%%%%%%%%%%%%%%%%%%%%%%%
In this article, we discuss CP property of approximately $\mu$-$\tau$ symmetric models 
satisfying $\det(M_\nu)=0$, whose theoretical foundation is supplied by the minimal seesaw model 
with two heavy right handed neutrinos. 
We estimate sizes of CP violating phases by using the general phase structure of neutrino mass 
matrix and by focusing on the rephasing invariance, whose existence in our formalism is discussed in 
Appendix \ref{sec:Appendix-rephasing}
. Some of results of the category (C1) are shared by 
our previous work \cite{Leptonic-CP-violation-induced-by-approximate-mu-tau}.  All CP phases are 
expressed in terms of flavor neutrino masses so that one can understand that how phases of 
flavor neutrino masses induce CP violating phases in $U_{PMNS}$.

%%%%%%%%%%%%%%%%%%%%%%%%%%%% Intro'Ü'Æ'ß %%%%%%%%%%%%%%%%%%%%%%%%%%%%
In the next section, we define the $\mu$-$\tau$ symmetry and explain two categories (C1) and (C2).
In Sec.\ref{sec:3}, we present various formulas to extract general property of the observed neutrino 
mixings and discuss how the condition of det($M_\nu$)=0 gives a massless neutrino to exclude 
the case of $m_2=0$. Detailed discussions to see the appearance of one massless neutrino 
are given by Appendix \ref{sec:Appendix-massless}.
In Sec.\ref{sec:4}, we include effects of the $\mu$-$\tau$ symmetry breaking to see neutrinos 
in the categories (C1) and (C2), which are used to construct neutrino mass textures.
In Sec.\ref{sec:5}, we argue how mass hierarchies are realized and show seven viable textures, 
where we estimate CP violating phases in each texture.
The last section is devoted to summary and discussions.

%%%%%%%%%%%%%%%%%%%%%%%%%%%%%%%%%%%%%%%%%%%%%%%%%%%%%%%%%%%%%%%%%%%%%%%%%%%%%%%%%%%%%%%%%%%%%%%%%%%%
%%%%%%%%%%%%%%%%%%%%%%%%%%%%%%%%%%%%%%%%%%%%%%%%%%%%%%%%%%%%%%%%%%%%%%%%%%%%%%%%%%%%%%%%%%%%%%%%%%%%
\section{\label{sec:2}$\mu$-$\tau$ symmetry}
%%%%%%%%%%%%%%%%%%Brief review of minimal seesaw model%%%%%%%%%%%%%%%%%%

The $\mu$-$\tau$ symmetry is the symmetry due to the invariance of the lagrangian, 
especially for the flavor neutrino mass term $M_\nu$, associated with the interchange 
of $\nu_\mu\leftrightarrow \nu_\tau$. 
We define the interchange as follows:
%%%%%%%%%%%%%%%%%%%%
\begin{equation}
\nu_\mu\leftrightarrow -\sigma\nu_\tau~(\sigma = \pm 1),
\label{Eq:mu-tau interchange}
\end{equation}
%%%%%%%%%%%%%%%%%%%%
where $\sigma$ will take care of the sign of $\sin\theta_{23}$ as parameterized by Eq.(\ref{Eq:PMNS matrix}). 
The phase $\gamma$ turns out to be of the $\mu$-$\tau$ 
symmetry breaking type. 
Our mass term $M_\nu$ is labeled by
%%%%%%%%%%%%%%%%%%%%
\begin{equation}
M_\nu   = \left( {\begin{array}{*{20}c}
   {M_{ee} } & {M_{e\mu } } & {M_{e\tau } }  \\
   {M_{e\mu } } & {M_{\mu \mu } } & {M_{\mu \tau } }  \\
   {M_{e\tau } } & {M_{\mu \tau } } & {M_{\tau \tau } }  \\
\end{array}} \right),
\label{Eq:flavor mass Mnu}
\end{equation}
%%%%%%%%%%%%%%%%%%%%
which is divided into $M^{(+)}_\nu$ and $M^{(-)}_\nu$ ($M_\nu=M^{(+)}_\nu$+$M^{(-)}_\nu$):
%%%%%%%%%%%%%%%%%%%%
\begin{equation}
M_\nu ^{\left(  +  \right)}  = \left( {\begin{array}{*{20}c}
   {M_{ee} } & {M_{e\mu }^{\left(  +  \right)} } & { - \sigma M_{e\mu }^{\left(  +  \right)} }  \\
   {M_{e\mu }^{\left(  +  \right)} } & {M_{\mu \mu }^{\left(  +  \right)} } & {M_{\mu \tau } }  \\
   { - \sigma M_{e\mu }^{\left(  +  \right)} } & {M_{\mu \tau } } & {M_{\mu \mu }^{\left(  +  
\right)} }  \\
\end{array}} \right),~
M_\nu ^{\left(  -  \right)}  = \left( {\begin{array}{*{20}c}
   0 & {M_{e\mu }^{\left(  -  \right)} } & {\sigma M_{e\mu }^{\left(  -  \right)} }  \\
   {M_{e\mu }^{\left(  -  \right)} } & {M_{\mu \mu }^{\left(  -  \right)} } & 0  \\
   {\sigma M_{e\mu }^{\left(  -  \right)} } & 0 & { - M_{\mu \mu }^{\left(  -  \right)} }  \\
\end{array}} \right).
\label{Eq:matrix elements of Mnu}
\end{equation}
%%%%%%%%%%%%%%%%%%%%
Under the interchange Eq.(\ref{Eq:mu-tau interchange}), $M^{(+)}_\nu$ is kept intact.
From this result, the superscripts $(+)$ and $(-)$ of $M_\nu$ are, respectively, so chosen to stand for 
 the $\mu$-$\tau$ symmetry preserving and breaking terms.

We find that $\sin\theta_{23}=\sigma/\sqrt{2}$ is determined by 
(0, $\sigma/\sqrt{2}$, $1/\sqrt{2}$)$^T$ as one of the eigenvectors associated 
with $M_\nu ^{\left(  +  \right)}$ if it is assigned to $\nu_3$. 
One may also assign it to $\nu_2$ giving 
$\sin\theta_{12}=0$, even to $\nu_1$ giving $\cos\theta_{12}=0$.  
Namely $U_{PMNS}$, respectively, takes the form of
%%%%%%%%%%%%%%%%%%%%
\begin{eqnarray}
&&\left( {\begin{array}{*{20}{c}}
    *  &  *  & 0  \\
    *  &  *  & {\frac{\sigma }{{\sqrt 2 }}}  \\
    *  &  *  & {\frac{1}{{\sqrt 2 }}}  \\
\end{array}} \right)~(\sin\theta_{13}=0),
\quad
\left( {\begin{array}{*{20}{c}}
    *  & 0 &  *   \\
    *  & {\frac{1}{{\sqrt 2 }}} &  *   \\
    *  & {\frac{\sigma }{{\sqrt 2 }}} &  *   \\
\end{array}} \right)~(\sin\theta_{12}=0),
\quad
\nonumber\\
&&\left( {\begin{array}{*{20}{c}}
   0 &  *  &  *   \\
   { - \frac{1}{{\sqrt 2 }}} &  *  &  *   \\
   { - \frac{\sigma }{{\sqrt 2 }}} &  *  &  *   \\
\end{array}} \right)~(\cos\theta_{12}=0).
\label{Eq:ThreeU_PMNS}
\end{eqnarray}
%%%%%%%%%%%%%%%%%%%%
There are in principle three possibilities for $\mu$-$\tau$ symmetric textures. 
However, 
for $\cos\theta_{12}=0$, after the $\mu$-$\tau$ symmetry is broken,
it can be shown that $\tan 2\theta_{12}={\mathcal {O}}$(1) 
but $\sin^2\theta_{12} > \cos^2 \theta_{12}$, which contradicts the data Eq.(\ref{Eq:experimental data}).
As a plausible choice, 
we obtain \cite{C1-C2}
%%%%%%%%%%%%%%%%%%%%
\begin{equation}
\sin\theta_{13}=0, \quad \sin\theta_{23}=\frac{\sigma}{\sqrt{2}},\quad \delta=\gamma=0,
\label{Eq:C1}
\end{equation}
%%%%%%%%%%%%%%%%%%%%
as a category (C1) or
%%%%%%%%%%%%%%%%%%%%
\begin{equation}
\sin\theta_{12}=0, \quad \sin\theta_{23}=-\frac{\sigma}{\sqrt{2}},\quad \delta+\rho=0, 
\quad \gamma=0,
\label{Eq:C2}
\end{equation}
%%%%%%%%%%%%%%%%%%%%
as a category (C2).  The phase $\rho$ is determined as $\rho$=arg($\sum^{\tau}_{i=e}M^\dagger_{ei}M_{i\mu}$) 
to be shown in Eq.(\ref{Eq:App rho delta from X-Y}), 
where $M_{ij}$ stands for 
an $i$-$j$ element of $M_\nu$ ($i,j$=$e,\mu,\tau$) defined in 
Eq.(\ref{Eq:flavor mass Mnu}). Both categories give no Dirac CP violation
signaled by $\sin\theta_{13}=0$ or by $\delta_{CP}(=\delta+\rho)=0$. In other words,
the Jarlskog invariant \cite{Jarlskog} vanishes.
In models without leptonic CP violation, 
both cases can produce experimentally allowed results \cite{C1-C2}, 
where the $\mu$-$\tau$ symmetry breaking is 
a must for the category (C2).  

%%%%%%%%%%%%%%%%%%%%%%%%%%%%%%%%%%%%%%%%%%%%%%%%%%%%%%%%%%%%%%%%%%%%%%%%%%%%%%%%%%%%%%%%%%%%%%%%%%%%
%%%%%%%%%%%%%%%%%%%%%%%%%%%%%%%%%%%%%%%%%%%%%%%%%%%%%%%%%%%%%%%%%%%%%%%%%%%%%%%%%%%%%%%%%%%%%%%%%%%%
\section{\label{sec:3}Various Relations of Masses and Mixings}
%%%%%%%%%%%%%%%%%%%%%%%%%%%%%%%%%%%%%%%%%%%%%%%%%%%%%%%%%%%%%%%%
To extract general property inherent to the observed neutrino mixings, 
we first derive various formulas expressed in terms of flavor neutrino masses
to evaluate neutrino masses, mixing angles and phases.  

%%%%%%%%%%%%%%%%%%%%%%%%%%%%%%%%%%%%%%%%%%%%%%%%%%%%%%%%%%%%%%%%
\subsection{\label{subsec:3-1}Useful Formulas}
%%%%%%%%%%%%%%%%%%%%%%%%%%%%%%%%%%%%%%%%%%%%%%%%%%%%%%%%%%%%%%%%
To get the Dirac CP phase $\delta$ as well as $\rho$ and $\gamma$, it is convenient to use 
${\rm\bf M} \left(\equiv M_{\nu}^{\dagger}M_{\nu}\right)$ parameterized by
%%%%%%%%%%%%%%%%%%%%
\begin{equation}
\textbf{ M }=
\left( {\begin{array}{*{20}c}
   A & B & C  \\
   {B^\ast  } & D & E  \\
   {C^\ast  } & {E^\ast  } & F  \\
\end{array}} \right),
\label{Eq:App M^daggerM}
\end{equation}
%%%%%%%%%%%%%%%%%%%%
where Majorana phases are removed.  From
%%%%%%%%%%%%%%%%%%%%
\begin{equation}
U^\dagger_{PMNS}\textbf{M}U_{PMNS}={\rm diag.}(m^2_1, m^2_2, m^2_3),
\label{Eq:App U^daggerM^daggerMU}
\end{equation}
%%%%%%%%%%%%%%%%%%%%
we obtain that
%%%%%%%%%%%%%%%%%%%%
\begin{eqnarray}
&&
 \tan 2\theta _{12} e^{i\rho }  = \frac{{2X}}{{\Lambda _2  - \Lambda _1 }},
 \quad
 \tan 2\theta _{13} e^{ - i\delta }  = \frac{{2Y}}{{\Lambda _3  - A}} 
\label{Eq:App MixingAngle12-13 from M^daggerM} \\ 
&&
 {\rm Re} \left( {e^{ - 2i\gamma } E} \right)\cos 2\theta _{23}  - \frac{{F - D}}{2}\sin 2
\theta _{23}  + i{\rm Im} \left( {e^{ - 2i\gamma } E} \right) =  - s_{13} e^{ - i\left( {
\rho  + \delta } \right)} \left( {e^{ - i\rho } X} \right)^\ast,
\nonumber\\
\label{Eq:App MixingAngle23 from M^daggerM} \\ 
&&
 m_1^2  = \frac{{\Lambda _1  + \Lambda _2 }}{2} - \frac{{e^{ - i\rho } X}}{{\sin 2\theta _{12} }},
\quad
 m_2^2  = \frac{{\Lambda _1  + \Lambda _2 }}{2} + \frac{{e^{ - i\rho } X}}{{\sin 2\theta _{12} }}, 
\quad
 m_3^2  = \frac{{c_{13}^2 \Lambda _3  - s_{13}^2 A}}{{c_{13}^2  - s_{13}^2 }},
\nonumber\\
 \label{Eq:App Mass from M^daggerM} \\ 
&&
 \Lambda _1  = \frac{{c_{13}^2 A - s_{13}^2 \Lambda _3 }}{{c_{13}^2  - s_{13}^2 }},
\quad
 \Lambda _2  = c_{23}^2 D + s_{23}^2 F - 2s_{23} c_{23} {\rm Re} \left( e^{ - 2i\gamma }E \right),
\nonumber\\
&&
 \Lambda _3  = s_{23}^2 D + c_{23}^2 F + 2s_{23} c_{23} {\rm Re} \left( e^{ - 2i\gamma }E \right), 
\label{Eq:App Lambda from M^daggerM} \\ 
&&
 X = \frac{{c_{23} e^{i\gamma } B - s_{23} e^{ - i{\gamma}} C}}{{c_{13} }},
 \quad
 Y = s_{23} e^{i\gamma } B + c_{23} e^{ - i\gamma} C, 
\label{Eq:App X Y from M^daggerM}
\end{eqnarray}
%%%%%%%%%%%%%%%%%%%%
Because $A$ and $\Lambda_{1,2,3}$ are real numbers, the phases $\rho$ and $\delta$ 
can be determined by
%%%%%%%%%%%%%%%%%%%%
\begin{equation}
\rho = {\rm arg}(X),
\quad
\delta = -{\rm arg}(Y).
\label{Eq:App rho delta from X-Y}
\end{equation}
%%%%%%%%%%%%%%%%%%%%
The remaining phase $\gamma$ is determined from Eq.(\ref{Eq:App MixingAngle23 from M^daggerM}).
It is further obtained that the size of $\left| X\right|$ should be suppressed to realize
the hierarchy of $\Delta m^2_\odot/\left|\Delta m^2_{atm}\right| \ll 1$. where
%%%%%%%%%%%%%%%%%%%%
\begin{equation}
\Delta m^2_\odot = 2\frac{{e^{ - i\rho } X}}{{\sin 2\theta _{12} }}~(>0). 
\label{Eq:App solar-squared from X}
\end{equation}
%%%%%%%%%%%%%%%%%%%%

To calculate the Majorana phases, we instead diagonalize $M_\nu$:
%%%%%%%%%%%%%%%%%%%%
\begin{equation}
U^T_{PMNS}M_\nu U_{PMNS}={\rm diag.}(m_1, m_2, m_3),
\label{Eq:App U^TMU}
\end{equation}
%%%%%%%%%%%%%%%%%%%%
and find that
%%%%%%%%%%%%%%%%%%%%
\begin{eqnarray}
&&
\tan 2\theta _{12}  =  \frac{{2x}}{{\lambda _2  - \lambda _1 }},
\quad
\tan 2\theta _{13} e^{i\rho }  = \frac{{2y}}{{\lambda _3 e^{i\delta }  - ae^{ - i\delta } }},
\nonumber\\ 
&&
 e\cos 2\theta _{23}  - \frac{{e^{ - 2i\gamma } f - e^{2i\gamma } d}}{2}\sin 2\theta _{23}
  =  - s_{13} e^{ - i\left( {\delta  + \rho } \right)} x,
\label{Eq:App MixingAngle from M}\\
&&
 m_1 e^{ - 2i\phi _1 }  = \frac{{\lambda _1  + \lambda _2 }}{2} - \frac{x}{{\sin 2\theta _{12} }},
 \quad
 m_2 e^{ - 2i\phi _2 }  = \frac{{\lambda _1  + \lambda _2 }}{2} + \frac{x}{{\sin 2\theta _{12} }},
\nonumber\\ 
&&
 m_3 e^{ - 2i\phi _3 }  = \frac{{c_{13}^2 \lambda _3  - s_{13}^2 e^{-2i\delta } a}}{{c_{13}
^2  - s_{13}^2 }},
 \label{Eq:App Mass from M} \\ 
&&
 \lambda _1  = e^{2i\rho } \frac{{c_{13}^2 a - s_{13}^2 e^{2i\delta } \lambda _3 }}{{c_{13}
^2  - s_{13}^2 }},
 \quad
 \lambda _2  = c_{23}^2 e^{2i\gamma } d + s_{23}^2 e^{ - 2i\gamma } f - 2s_{23} c_{23} e,
\nonumber\\ 
&&
 \lambda _3  = s_{23}^2 e^{2i\gamma } d + c_{23}^2 e^{ - 2i\gamma } f + 2s_{23} c_{23} e, 
\label{Eq:App lambda from M} \\ 
&&
x = e^{i\rho } \frac{{c_{23} e^{i\gamma } b - s_{23} e^{ - i\gamma } c}}{{c_{13} }},
\quad
y = e^{i\rho } \left( {s_{23} e^{i\gamma } b + c_{23} e^{ - i\gamma } } \right).
\label{Eq:App X Y from M}
\end{eqnarray}
%%%%%%%%%%%%%%%%%%%%
If $x$ is suppressed, $\lambda_1\approx \lambda_2$ to give a sizable $\theta_{12}$.
The Majorana phases $\phi_{1,2}$ become the similar order for the inverted mass hierarchy 
with $m_1\approx m_2$.  In this case, for det($M_\nu$)=0 giving $m_3=0$, 
Majorana CP violation is generically small.

%%%%%%%%%%%%%%%%%%%%%%%%%%%%%%%%%%%%%%%%%%%%%%%%%%%%%%%%%%%%%%%%
\subsection{\label{subsec:3-2}det(M$_\nu$)=0}
%%%%%%%%%%%%%%%%%%%%%%%%%%%%%%%%%%%%%%%%%%%%%%%%%%%%%%%%%%%%%%%%
Let us next see how the condition of $\det(M_\nu)=0$ provides a massless neutrino when our 
formulas of $m_{1,2,3}$ are used. 
Since the violation of the $\mu$-$\tau$ symmetry is tiny, it does not significantly affect the sizes of 
the neutrino masses evaluated in the $\mu$-$\tau$ symmetric limit although it affects the size of $\theta_{12}$ 
for the category (C2).  The obtained results are to be used to construct textures either with $m_1=0$ or $m_3=0$
but not with $m_2=0$.

If we parameterize $M_\nu$ as follows:
%%%%%%%%%%%%%%%%%%%%
\begin{equation}
M_\nu   = \left( {\begin{array}{*{20}c}
   a & b & c  \\
   b & d & e  \\
   c & e & f  \\
\end{array}} \right),
\label{Eq:App M_abcdef}
\end{equation}
%%%%%%%%%%%%%%%%%%%%
$\det(M_\nu)=0$ is translated into:
%%%%%%%%%%%%%%%%%%%%
\begin{equation}
e = \frac{{bc + s\sigma \sqrt {\left( {bc} \right)^2  - a\left( {b^2 f + c^2 d - adf}
 \right)} }}{a},
\label{Eq:e from det(Mnu)}
\end{equation}
%%%%%%%%%%%%%%%%%%%%
where $s=\pm 1$. The factor $\sigma$ may not be needed in Eq.(\ref{Eq:e from det(Mnu)}); 
however, it intends to take care of $\sigma$ from $c=-\sigma b$ in the $\mu$-$\tau$ symmetric limit 
and it is merely our matter of convention.
In the $\mu$-$\tau$ symmetric limit, Eq.(\ref{Eq:e from det(Mnu)}) turns out to be
%%%%%%%%%%%%%%%%%%%%
\begin{eqnarray}
 \sigma e &=&  - d,
\label{Eq:e from mu-tau symmetric det(Mnu) 1}
\end{eqnarray}
%%%%%%%%%%%%%%%%%%%%
for$\sqrt {\left( {b^2  - ad} \right)^2 }  =  s \left( {b^2  - ad} \right)$, and
%%%%%%%%%%%%%%%%%%%%
\begin{eqnarray}
 \sigma e &=&  d-\frac{2b^2}{a},
\label{Eq:e from mu-tau symmetric det(Mnu) 2}
\end{eqnarray}
%%%%%%%%%%%%%%%%%%%%
for $\sqrt {\left( {b^2  - ad} \right)^2 }  = - s \left( b^2  - ad \right)$. 
One can also confirm that the condition of $\det(M_\nu)=0$ evaluated
up to the first order of the $\mu$-$\tau$ symmetry breaking
coincides with that of $\det(M_\nu)=0$ in 
the $\mu$-$\tau$ symmetric limit.

The appearance of a massless neutrino is described in Appendix \ref{sec:Appendix-massless} using
Eq.(\ref{Eq:App Mass from M}).  The results are summarized as follows:
%%%%%%%%%%%%%%%%%%%%
\begin{itemize}
\item either $m_1=0$ or $m_2=0$ from Eq.(\ref{Eq:e from mu-tau symmetric det(Mnu) 2}) 
%%%%%%%%%%%%%%%%%%%%
\begin{eqnarray}
&& m_1=0~{\rm if}~ \sqrt {z^2}  = z,
\label{Eq:C1 m1=0}\\
&& m_2=0~{\rm if}~ \sqrt {z^2}  = -z,
\label{Eq:C1 m2=0}
\end{eqnarray}
%%%%%%%%%%%%%%%%%%%%
where $z$ = $d  - \sigma e  + e^{2i\rho } a$, 
\item  $m_3=0$ from Eq.(\ref{Eq:e from mu-tau symmetric det(Mnu) 1}),
\end{itemize}
%%%%%%%%%%%%%%%%%%%%
for the category (C1), where $\sin\theta_{23}= \sigma/\sqrt{2}$ and 
$\sin\theta_{13}=0$ in the $\mu$-$\tau$ symmetric limit, and
%%%%%%%%%%%%%%%%%%%%
\begin{itemize}
\item either $m_1=0$ or $m_3=0$ from Eq.(\ref{Eq:e from mu-tau symmetric det(Mnu) 2}) 
%%%%%%%%%%%%%%%%%%%%
\begin{eqnarray}
&& m_1=0~{\rm if}~ \sqrt {z^2}  = kz,
\label{Eq:C2 m1=0}\\
&& m_3=0~{\rm if}~ \sqrt {z^2}  = -kz,
\label{Eq:C2 m3=0}
\end{eqnarray}
%%%%%%%%%%%%%%%%%%%%
where $k(=\pm 1$) takes care of the sign of $\cos 2\theta _{13}$ 
and $z$ = $\left( {d  - \sigma e } \right)e^{i\delta }  + a e^{ - i\delta } $, 
%%%%%%%%%%%%%%%%%%%%
\item $m_2=0$ from Eq.(\ref{Eq:e from mu-tau symmetric det(Mnu) 1})
\end{itemize}
%%%%%%%%%%%%%%%%%%%%
for the category (C2), where $\sin\theta_{23}= -\sigma/\sqrt{2}$ and 
$\sin\theta_{12}=0$ in the $\mu$-$\tau$ symmetric limit.  In any cases, the massless $\nu_2$ should 
not be realized by textures.

%%%%%%%%%%%%%%%%%%%%%%%%%%%%%%%%%%%%%%%%%%%%%%%%%%%%%%%%%%%%%%%%%%%%%%%%%%%%%%%%%%%%%%%%%%%%%%%%%%%%
%%%%%%%%%%%%%%%%%%%%%%%%%%%%%%%%%%%%%%%%%%%%%%%%%%%%%%%%%%%%%%%%%%%%%%%%%%%%%%%%%%%%%%%%%%%%%%%%%%%%
\section{\label{sec:4}Effect of $\mu$-$\tau$ Symmetry Breaking}
%%%%%%%%%%%%%%%%%%%%%%%%%%%%%%%%%%%%%%%%%%%%%%%%%%%%%%%%%%%%%%%%%%%%%%%%%%%%%%%%%%%%%%%%%%%%%%%%
To specify phase structure of $M_\nu$, let us first count phases present in $M_\nu$.  There are six 
complex numbers in $M_\nu$.  Since three 
phases are removed by the rephasing, among the remaining three phases, one phase 
can be determined by $\det(M_\nu)=0$. We are left with two phases, which are taken to be the phases
associated with $M_{e\mu}$ and $M_{e\tau}$. For the present discussions, these phases are 
denoted by $\alpha$ associated with $M^{(+)}_{e\mu}$ and by $\beta$ associated with 
$M^{(-)}_{e\mu}$ in place of $M_{e\mu}$ and $M_{e\tau}$ for the sake of convenience. Any other 
choices give the same results of CP violation because of the rephasing invariance as shown 
in Appendix \ref{sec:Appendix-rephasing}. Thus, our results do not depend on our specific choice of phases in $M_\nu$ 
and will cover leptonic CP properties in models with $\det(M_\nu)=0$, where the charged lepton masses 
are necessarily taken to diagonal \cite{Non-diagonal-charged-lepton-mass-matrix}.

%%%%%%%%%%%%%%%%%%%%%%%%%%%%%%%%%%%%%%%%%%%%%%%%%%%%%%%%%%%%%%%%
\subsection{\label{subsec:4-1}Parameterization of $\mu$-$\tau$ Symmetry Breaking}
%%%%%%%%%%%%%%%%%%%%%%%%%%%%%%%%%%%%%%%%%%%%%%%%%%%%%%%%%%%%%%%%
To describe the $\mu$-$\tau$ symmetry breaking flavor neutrino masses,
we parameterize
%%%%%%%%%%%%%%%%%%%%
\begin{equation}
M^{(+)}_\nu=
\left( {\begin{array}{*{20}c}
   a_0 & e^{i\alpha}b_0 & -\sigma e^{i\alpha}b_0  \\
   e^{i\alpha}b_0 & d_0 & e_0  \\
   -\sigma e^{i\alpha}b_0 & e_0 & d_0  \\
\end{array}} \right)
\quad
M^{(-)}_\nu=
\varepsilon\left( {\begin{array}{*{20}c}
   0 & e^{i\beta}b^\prime_0 & \sigma e^{i\beta}b^\prime_0  \\
   e^{i\beta}b^\prime_0 & d^\prime_0 & 0  \\
   \sigma e^{i\beta}b^\prime_0 & 0 & -d^\prime_0  \\
\end{array}} \right),
\label{Eq:App M_abcdef-breaking}
\end{equation}
%%%%%%%%%%%%%%%%%%%%
for Eq.(\ref{Eq:matrix elements of Mnu}) and
%%%%%%%%%%%%%%%%%%%%
\begin{equation}
{\rm\bf M}^{(+)}  = \left( \begin{array}{*{20}c}
   A & B_+ & - \sigma B_+  \\
   B^\ast_+ & D_+ & E_+   \\
   - \sigma B^\ast_+ & E_+ & D_+\\
\end{array} \right),
\quad
{\rm\bf M}^{(-)}  = \left( \begin{array}{*{20}c}
   0 & B_- & \sigma B_-  \\
   B^\ast_- & D_- & iE_-  \\
   \sigma B^\ast_- & -iE_- & - D_-\\
\end{array} \right),
\label{Eq:App M^daggerM-breaking}
\end{equation}
%%%%%%%%%%%%%%%%%%%%
where ${\rm\bf M}\equiv {\rm\bf M}^{(+)} +{\rm\bf M}^{(-)}$.
We show results valid up to the terms of ${\mathcal O}(\varepsilon)$.\footnote{It should be 
noted that the smallness of $\Delta m^2_\odot/\vert\Delta m^2_{atm}\vert$ is naturally 
${\mathcal O}(\varepsilon)$ in the category (C2) but 
the smallness of $\sin\theta_{13}$ is also implicitly assumed to be consistent with 
the experimental observation.  It is realized by
the smallness of $b_0$ contained in $y$ (See Eq.(\ref{Eq:App masses and mixing angles M})
 for $\kappa=-1$).}  To do so, we 
parameterize $\theta_{23}$ as follows:
%%%%%%%%%%%%%%%%%%%%
\begin{equation}
\cos\theta_{23}  = \frac{{1 + \Delta }}{{\sqrt {2\left( {1 + \Delta ^2 } \right)} }},
~
\sin\theta_{23}  = \kappa \sigma \frac{{1 - \Delta }}{{\sqrt {2\left( {1 + \Delta ^2 } \right)} }},
\label{Eq:App cos23-sin23}
\end{equation}
%%%%%%%%%%%%%%%%%%%%
where $\vert\Delta\vert (\ll 1)$ is responsible for the effect of ${\mathcal O}(\varepsilon)$.  
The parameter $\kappa(=\pm 1)$ takes care of the difference of the category: $\kappa=1$ for 
the category (C1) and $\kappa=-1$ for the category (C2).  

The results are given by
%%%%%%%%%%%%%%%%%%%%
\begin{eqnarray}
 &&
 A \approx \left| {a_0 } \right|^2  + 2\left| {b_0 } \right|^2 , 
\quad
 B_ +   \approx b_0 \left( {e^{i\alpha } a_0  + e^{ - i\alpha } \left( {d_0  - \sigma e_0 }
 \right)} \right), 
 \nonumber\\
 &&
 B_ -   \approx 
 \varepsilon\left[ {\left( {a_0 e^{i\beta }  + e^{ - i\beta } \left( {d_0  + \sigma e_0 } \right)} \right)b^\prime_0  + e^{ - i\alpha } b_0 d^\prime_0 } \right],
 \nonumber\\
&&
 D_ +   \approx \left| {b_0 } \right|^2  + \left| {d_0 } \right|^2  + \left| {e_0 } \right|^2 , 
\quad
 D_ -   = 2\varepsilon \left[ {b_0 b^\prime_0 \cos \left( {\alpha  - \beta } \right) + d_0 d^\prime_0 }
 \right], 
 \nonumber\\
&&
 E_ +   \approx 2d_0 {\rm Re} \left( {e_0 } \right) - 
        \sigma \left| {b_0 } \right|^2 ,{\rm{ }}
\quad
 E_ -   = 2\varepsilon \left[ {d^\prime_0 {\rm Im} \left( {e_0 } \right) - 
        \sigma b_0 b^\prime_0 \sin \left( {\alpha  - \beta } \right)} \right]
 \nonumber\\
 &&
 \Lambda _1  \approx A, 
\quad
 \Lambda _2  \approx 
 \left\{ \begin{array}{l}
 2b_0^2  + \left| {d_0  - \sigma e_0 } \right|^2 ~\left( {\kappa  = 1} \right) \\ 
 \left| {d_0  + \sigma e_0 } \right|^2 ~\left( {\kappa  =  - 1} \right)\\ 
 \end{array} \right.,
 \nonumber\\
 &&
  \Lambda _3  \approx
  \left\{ \begin{array}{l}
 \left| {d_0  + \sigma e_0 } \right|^2 ~\left( {\kappa  = 1} \right) \\ 
 2b_0^2  + \left| {d_0  - \sigma e_0 } \right|^2 ~\left( {\kappa  =  - 1} \right) \\ 
 \end{array} \right.,
 \nonumber\\
&&
 X \approx 
\left\{ \begin{array}{l}
 \sqrt 2 B_ +  ~\left( {\kappa  = 1} \right) \\ 
 \sqrt 2 \left( {B_ -   + \left( {\Delta  + i\gamma } \right)B_ +  } \right)~\left( {\kappa  
=  - 1} \right) \\ 
 \end{array} \right., 
 \nonumber\\
 &&
 Y \approx 
\left\{ \begin{array}{l}
 \sqrt 2 \sigma \left( {B_ -   - \left( {\Delta  - i\gamma } \right)B_ +  } \right)~\left( 
{\kappa  = 1} \right) \\ 
 - \sqrt 2 \sigma B_ +  ~\left( {\kappa  =  - 1} \right) \\ 
 \end{array} \right.,
 \nonumber\\
&&
 \Delta  \approx -\frac{{\kappa \sigma D_ -    + s_{13} {\rm Re} \left( {e^{ - i\delta } X^ 
*  } \right)}}{{2E_ +  }}, 
\quad
 \gamma  \approx \frac{{E_ -   + s_{13} {\rm Im} \left( {e^{ - i\delta } X^ *  } \right)}}
{{2E_ +  }}, 
\label{Eq:App masses and mixing angles MdaggerM}
\end{eqnarray}
%%%%%%%%%%%%%%%%%%%%
from which the mixing angles in Eq.(\ref{Eq:App MixingAngle from M}) and the Dirac phases 
in Eq.(\ref{Eq:App rho delta from X-Y}) can be estimated, and
%%%%%%%%%%%%%%%%%%%%
\begin{eqnarray}
 &&
 \lambda _1  \approx e^{2i\rho } a_0 ,~\lambda _2  \approx d_0  - \kappa \sigma e_0 ,~\lambda 
_3  \approx d_0  + \kappa \sigma e_0, 
 \nonumber\\ 
 &&
 x \approx \left\{ \begin{array}{l}
 2e^{i\left( {\rho  + \alpha } \right)} b_0 ~\left( {\kappa  = 1} \right) \\ 
 2e^{i\rho } \left( {\left( {\Delta  + i\gamma } \right)e^{i\alpha } b_0  + \varepsilon e^
{i\beta } b^\prime_0 } \right)~\left( {\kappa  = -1} \right) \\ 
 \end{array} \right.,
 \nonumber\\ 
 &&
 y \approx \left\{ \begin{array}{l}
 2\sigma e^{i\rho } \left( {\left( { - \Delta  + i\gamma } \right)e^{i\alpha } b_0  + 
\varepsilon e^{i\beta } b^\prime_0 } \right)~\left( {\kappa  = 1} \right) \\ 
 -2\sigma e^{i\left( {\rho  + \alpha } \right)} b_0 ~\left( {\kappa  = -1} \right) \\ 
 \end{array} \right.,
 \nonumber\\ 
 &&
 m_1 e^{ - 2i\varphi _1 }  \approx \frac{{e^{2i\rho } a_0  + d_0  - \kappa \sigma e_0 }}{2}
 - \frac{x}{{\sin 2\theta _{12} }},
 \nonumber\\ 
 &&
 m_2 e^{ - 2i\varphi _2 }  \approx \frac{{e^{2i\rho } a_0  + d_0  - \kappa \sigma e_0 }}{2}
 + \frac{x}{{\sin 2\theta _{12} }},
 \nonumber\\ 
 &&
 m_3 e^{ - 2i\varphi _3 }  \approx d_0  + \kappa \sigma e_0, 
\label{Eq:App masses and mixing angles M}
\end{eqnarray}
%%%%%%%%%%%%%%%%%%%%
from which Majorana phases can be estimated. 

%%%%%%%%%%%%%%%%%%%%%%%%%%%%%%%%%%%%%%%%%%%%%%%%%%%%%%%%%%%%%%%%
\subsection{\label{subsec:4} Masses, Mixing Angles and Phases}
%%%%%%%%%%%%%%%%%%%%%%%%%%%%%%%%%%%%%%%%%%%%%%%%%%%%%%%%%%%%%%%%
Before we give explicit textures,  
we calculate masses, mixing angles and phases in terms of the mass parameters of 
Eq.(\ref{Eq:App M_abcdef-breaking}) from relations found in the 
previous subsection for each category. Explicit forms of textures can readily be obtained once 
we give mass parameters in Eq.(\ref{Eq:App M_abcdef-breaking}), which are 
taken to realize the normal mass hierarchy or the inverted mass hierarchy.

%%%%%%%%%%%%%%%%%%%%%%%%%%%%%%%%%%%%%%%%%%%%%%%%%%%%%%%%%%%%%%%%
\subsubsection{Category (C1)}
%%%%%%%%%%%%%%%%%%%%%%%%%%%%%%%%%%%%%%%%%%%%%%%%%%%%%%%%%%%%%%%%
Mixing angles and Dirac phases can be 
estimated in terms of Eq.(\ref{Eq:App masses and mixing angles MdaggerM}) as
%%%%%%%%%%%%%%%%%%%%
\begin{eqnarray}
 \tan 2\theta _{12} e^{i\rho } &\approx& \frac{{2X}}{{\left( {\left| {d_0  - \sigma e_0 } \right| + \left| {a_0 } \right|} \right)\left( {\left| {d_0  - \sigma e_0 } \right| - \left| {a_0 } \right|} \right)}},
 \nonumber\\
 \tan 2\theta _{13} e^{ - i\delta } &\approx& \frac{{2Y}}{{\left( {\left| {d_0  + \sigma e_0 } \right| + \sqrt {\left| {a_0 } \right|^2  + 2\left| {b_0 } \right|^2 } } \right)\left( {\left| {d_0  + \sigma e_0 } \right| - \sqrt {\left| {a_0 } \right|^2  + 2\left| {b_0 } \right|^2 } } \right)}},
 \nonumber\\
 \cos 2\theta _{23} &\approx& 2\Delta,
\label{Eq:App C1 mixing angles}
\end{eqnarray}
%%%%%%%%%%%%%%%%%%%%
where
%%%%%%%%%%%%%%%%%%%%
\begin{eqnarray}
 X &\approx& \sqrt 2 b_0 \left( {e^{i\alpha } a_0  + e^{ - i\alpha } \left( {d_0  - \sigma e_0 } \right)} \right),
 \nonumber\\
 Y &\approx& \sqrt 2 \sigma \left[ \begin{array}{l}
 \varepsilon \left( {\left( {{a_0}{e^{i\beta }} + {e^{ - i\beta }}\left( {{d_0} + \sigma {e_0}} \right)} \right)b_0^\prime  + {e^{ - i\alpha }}{b_0}d_0^\prime } \right) \\ 
 \quad - \left( {\Delta  - i\gamma } \right)\left( {{b_0}\left( {{e^{i\alpha }}{a_0} + {e^{ - i\alpha }}\left( {{d_0} - \sigma {e_0}} \right)} \right)} \right) \\ 
 \end{array} \right],
 \nonumber\\
 \Delta &\approx& 
-\frac{{2\sigma \varepsilon \left[ {b_0 b^\prime_0 \cos \left( {\alpha  - \beta } \right) + d_0 d^\prime_0 } \right] - s_{13} {\rm Re} \left( {e^{ - i\delta } X^ *  } \right)}}{{2\left( {2d_0 {\rm Re} \left( {e_0 } \right) - \sigma \left| {b_0 } \right|^2 } \right)}},
 \nonumber\\
\gamma  &\approx& \frac{{2\varepsilon \left[ {d^\prime_0 {\rm Im} \left( {e_0 } \right) - \sigma b_0 b^\prime_0 \sin \left( {\alpha  - \beta } \right)} \right] + s_{13} {\rm Im} \left( {e^{ - i\delta } X^ *  } \right)}}{{2\left( {2d_0 {\rm Re} \left( {e_0 } \right) - \sigma \left| {b_0 } \right|^2 } \right)}},
\label{Eq:App C1 X-Y-Delta}
\end{eqnarray}
%%%%%%%%%%%%%%%%%%%%
leading to
%%%%%%%%%%%%%%%%%%%%
\begin{equation}
\tan \rho  \approx \frac{{a_0  - \left(d_0-\sigma e_0 \right)}}{{a_0  + \left(d_0-\sigma e_0 \right) }}\tan \alpha,
\label{Eq:App C1 frho-alpha}
\end{equation}
%%%%%%%%%%%%%%%%%%%%
from $X$.  Since $\tan 2\theta_{13}$ involves $s_{13}$ via $\Delta$ and $\gamma$ in its right-hand side, 
one can further solve $\tan 2\theta_{13}$ to give
%%%%%%%%%%%%%%%%%%%%
\begin{eqnarray}
s_{13} e^{ - i\delta }  \approx \frac{{\sqrt 2 \sigma }}{{\Lambda _3  - A}}\frac{{B_ -   - B_ +  \frac{{D_ -  \kappa \sigma  - iE_ -  }}{{2E_ +  }}}}{{1 - \frac{{\sqrt 2 \sigma }}{{\Lambda _3  - A}}B_ +  \frac{{X^ *  }}{{2E_ +  }}}},
\label{Eq:App C1 s13}
\end{eqnarray}
%%%%%%%%%%%%%%%%%%%%
where $B_\pm$ and $E_\pm$ are given by Eq.(\ref{Eq:App masses and mixing angles MdaggerM}), 
which gives $\vert s_{13}\vert={\mathcal O}(\varepsilon)$.
Neutrino masses and Majorana phases are 
estimated to be:
%%%%%%%%%%%%%%%%%%%%
\begin{eqnarray}
 m_1 e^{ - 2i\varphi _1 }  &=& 0, \quad
 \nonumber\\
 m_2 e^{ - 2i\varphi _2 }  &\approx& e^{2i\rho } a_0  + d_0  - \sigma e_0, 
 \nonumber\\
 m_3 e^{ - 2i\varphi _3 }  &\approx& d_0  + \sigma e_0,
\label{Eq:App C1 normal masses and Majorana phases}
\end{eqnarray}
%%%%%%%%%%%%%%%%%%%%
for the normal mass hierarchy, and
%%%%%%%%%%%%%%%%%%%%
\begin{eqnarray}
 m_1 e^{ - 2i\varphi _1 } &\approx& \frac{{e^{2i\rho } a_0  + d_0  - \sigma e_0 }}{2} - \frac{x}{{\sin 2\theta _{12} }},
 \nonumber\\
 m_2 e^{ - 2i\varphi _2 } &\approx& \frac{{e^{2i\rho } a_0  + d_0  - \sigma e_0 }}{2} + \frac{x}{{\sin 2\theta _{12} }},
 \nonumber\\
 m_3 e^{ - 2i\varphi _3 } (=0) &\approx& d_0+\sigma e_0,
\label{Eq:App C1 inverted masses and Majorana phases}
\end{eqnarray}
%%%%%%%%%%%%%%%%%%%%
for the inverted mass hierarchy, where
%%%%%%%%%%%%%%%%%%%%
\begin{eqnarray}
 x \approx 
 2e^{i\left( {\rho  + \alpha } \right)} b_0.
\label{Eq:App C1 x}
\end{eqnarray}
%%%%%%%%%%%%%%%%%%%%
 
%%%%%%%%%%%%%%%%%%%%%%%%%%%%%%%%%%%%%%%%%%%%%%%%%%%%%%%%%%%%%%%%
\subsubsection{Category (C2)}
In this category, mixing angles and Dirac phases can be 
estimated in terms of Eq.(\ref{Eq:App masses and mixing angles MdaggerM}) as
%%%%%%%%%%%%%%%%%%%%
\begin{eqnarray}
 \tan 2\theta _{12} e^{i\rho } &\approx& \frac{{2X}}{{\left( {\left| {d_0  + \sigma e_0 } \right| - \sqrt {\left| {a_0 } \right|^2 + 2\left| {b_0 } \right|^2 } } \right)\left( {\left| {d_0  + \sigma e_0 } \right| + \sqrt {\left| {a_0 } \right|^2 + 2\left| {b_0 } \right|^2 } } \right)}},
 \nonumber\\
 \tan 2\theta _{13} e^{ - i\delta } &\approx& \frac{{2Y}}{{\left( {\left| {d_0  - \sigma e_0 } \right| - \left| {a_0 } \right|} \right)\left( {\left| {d_0  - \sigma e_0 } \right| + \left| {a_0 } \right|} \right)}},
 \nonumber\\
 \cos 2\theta _{23} &\approx& 2\Delta,
\label{Eq:App C2 mixing angles}
\end{eqnarray}
%%%%%%%%%%%%%%%%%%%%
where
%%%%%%%%%%%%%%%%%%%%
\begin{eqnarray}
X &\approx& \sqrt 2 \left[ \begin{array}{l}
 \varepsilon \left( {\left( {{a_0}{e^{i\beta }} + {e^{ - i\beta }}\left( {{d_0} + \sigma {e_0}} \right)} \right)b_0^\prime  + {e^{ - i\alpha }}{b_0}d_0^\prime } \right) \\ 
 \quad + \left( {\Delta  + i\gamma } \right){b_0}\left( {{e^{i\alpha }}{a_0} + {e^{ - i\alpha }}\left( {{d_0} - \sigma {e_0}} \right)} \right) \\ 
 \end{array} \right], 
 \nonumber\\
 Y &\approx&  - \sqrt 2 \sigma b_0 \left( {e^{i\alpha } a_0  + e^{ - i\alpha } \left( {d_0  - \sigma e_0 } \right)} \right),
 \nonumber\\
 \Delta &\approx&  
  \frac{{2\sigma \varepsilon \left[ {b_0 b^\prime_0 \cos \left( {\alpha  - \beta } \right) + d_0 d^\prime_0 } \right] + s_{13} {\rm Re} \left( {e^{ - i\delta } X^ *  } \right)}}{{2\left( {2d_0 {\rm Re} \left( {e_0 } \right) - \sigma \left| {b_0 } \right|^2 } \right)}},
\label{Eq:App C2 X-Y-Delta}
\end{eqnarray}
%%%%%%%%%%%%%%%%%%%%
with $\gamma$ replaced by $-\gamma$ in Eq.(\ref{Eq:App C1 X-Y-Delta}),
leading to
%%%%%%%%%%%%%%%%%%%%
\begin{equation}
\tan \delta  \approx \frac{{a_0  - \left(d_0-\sigma e_0 \right)}}{{a_0  + \left(d_0-\sigma e_0 \right) }}\tan \alpha,
\label{Eq:App C2 delta-alpha}
\end{equation}
%%%%%%%%%%%%%%%%%%%%
from $Y$.  Neutrinos exhibit the inverted mass hierarchy (as to be discussed in Sec.\ref{subsec:C2)}), 
where masses and Majorana phases are given by
%%%%%%%%%%%%%%%%%%%%
\begin{eqnarray}
 m_1 e^{ - 2i\varphi _1 } &\approx& \frac{{e^{2i\rho } a_0  + d_0  + \sigma e_0 }}{2} - \frac{x}{{\sin 2\theta _{12} }},
 \nonumber\\
 m_2 e^{ - 2i\varphi _2 } &\approx& \frac{{e^{2i\rho } a_0  + d_0  + \sigma e_0 }}{2} + \frac{x}{{\sin 2\theta _{12} }},
 \nonumber\\
 m_3 e^{ - 2i\varphi _3 } (=0) &\approx&  d_0-\sigma e_0,
\label{Eq:App C2 inverted masses and Majorana phases}
\end{eqnarray}
%%%%%%%%%%%%%%%%%%%%
where
d%%%%%%%%%%%%%%%%%%%%
\begin{eqnarray}
 x \approx 
 2e^{i\rho } \left(\left( {\Delta  + i\gamma } \right)e^{i\alpha } b_0  + \varepsilon e^
{i\beta } b^\prime_0 \right).
\label{Eq:App C2 x}
\end{eqnarray}
%%%%%%%%%%%%%%%%%%%%

One has to finetune parameters to yield a tiny quantity $\eta$ to be used in our textures.
This finetuning provides the smallness of
%%%%%%%%%%%%%%%%%%%%
\begin{enumerate}
\item $\Delta m^2_\odot$ in Eq.(\ref{Eq:App solar-squared from X}) requiring $\left| X \right|\approx 0$ 
in (\ref{Eq:App C1 X-Y-Delta}) that leads to either
$b_0\approx 0$ or $\left| {e^{i\alpha } a_0  + e^{ - i\alpha } \left( {d_0  - \sigma e_0 } \right)} \right|\approx 0$ 
 for the category (C1) and
\item $\sin\theta_{13}$ requiring $b_0\approx 0$ in Eq.(\ref{Eq:App C2 X-Y-Delta})
for the category (C2). 
\end{enumerate}
%%%%%%%%%%%%%%%%%%%%
All textures are so parameterized to satisfy these conditions.

%%%%%%%%%%%%%%%%%%%%%%%%%%%%%%%%%%%%%%%%%%%%%%%%%%%%%%%%%%%%%%%%%%%%%%%%%%%%%%%%%%%%%%%%%%%%%%%%%%%%
%%%%%%%%%%%%%%%%%%%%%%%%%%%%%%%%%%%%%%%%%%%%%%%%%%%%%%%%%%%%%%%%%%%%%%%%%%%%%%%%%%%%%%%%%%%%%%%%%%%%
\section{\label{sec:5}Mass Hierarchies and CP Violation}
%%%%%%%%%%%%%%%%%%%%%%%%%%%%%%%%%%%%%%%%%%%%%%%%%%%%%%%%%%%%%%%%
In this section, we select 
specific mass matrices to see how CP violations depend on phases of flavor neutrino masses. Some of 
these textures are those extrapolated from textures without no phases, which
 have been shown to consistently describe the current neutrino oscillations \cite{C1-C2}. 
Other textures are those having nontrivial forms that cannot be extrapolated 
from the textures without phases. Such nontrivial textures arise in the inverted mass hierarchy.  

%%%%%%%%%%%%%%%%%%%%%%%%%%%%%%%%%%%%%%%%%%%%%%%%%%%%%%%%%%%%%%%%
\subsection{Textures and Effect of Phases in the Inverted Mass Hierarchy}
%%%%%%%%%%%%%%%%%%%%%%%%%%%%%%%%%%%%%%%%%%%%%%%%%%%%%%%%%%%%%%%%
In the inverted mass hierarchy, for flavor neutrino masses expressed in terms of Eq.(\ref{Eq:App M_abcdef-breaking}),
we require that $\Delta m^2_\odot/\vert\Delta m^2_{atm}\vert \ll 1$ be satisfied and obtain 
from Eq.(\ref{Eq:App masses and mixing angles M})
that
%%%%%%%%%%%%%%%%%%%%
\begin{eqnarray}
&&
\left( {a_0  + 2d_0 } \right){\rm Re} \left( {e^{ - i\rho } x} \right)\cos \rho  + \left( {a_0  - 2d_0 } \right){\rm Im} \left( {e^{ - i\rho } x} \right)\sin \rho 
\approx 0,
%\label{Eq:Delta_sun/Delta_atm for complex m1=m2}
\label{Eq:Delta_sun/Delta_atm for complex m1=-m2}
\end{eqnarray}
%%%%%%%%%%%%%%%%%%%%
where we have used $d_0  + \kappa \sigma e_0\approx 0$ ($\kappa=1$ for the category (C1) and $\kappa = -1$ for the category (C2)) 
from $m_3=0$.  There are two solutions:
%%%%%%%%%%%%%%%%%%%%
\begin{enumerate}
\item %'±'±'É•¶'ð'¢'ê'é
$a_0  + 2d_0 \approx 0$ and ${\rm Im} \left( {e^{ - i\rho } x} \right)\sin \rho\approx 0$,
\item 
$a_0  - 2d_0 \approx 0$ and ${\rm Re} \left( {e^{ - i\rho } x} \right)\cos \rho\approx 0$.
\end{enumerate}
%%%%%%%%%%%%%%%%%%%%
The simplest case is $x\approx 0$, which is the case with $m_1\approx m_2$. If $x$ is not suppressed, 
textures with $m_1\approx -m_2$ are realized \cite{m1=-m2} and the specific interplay between $\rho$ and $x$ may give the 
necessary suppression.

For the category (C1), if $x\approx 0$, since $x\propto b_0$ 
as in Eq.(\ref{Eq:App masses and mixing angles M}), $b_0$ 
should be suppressed while 
%%%%%%%%%%%%%%%%%%%%
\begin{equation}
\left|{\cos \rho \left( {a_0  + 2d_0 } \right) + i\sin \rho \left( {a_0  - 2d_0 } \right)}\right|\neq 0,
\label{Eq:C1 scale of neutrino masses}
\end{equation}
%%%%%%%%%%%%%%%%%%%%
should be maintained to control the scale of neutrino masses as can 
be seen from Eq.(\ref{Eq:App C1 inverted masses and Majorana phases}).
 Because $\tan 2\theta_{12}$ in Eq.(\ref{Eq:App C1 mixing angles}) contains $b_0$ 
in the numerator, the denominator should have a factor to cancel the smallness of $b_0$.  Using 
Eq.(\ref{Eq:App C1 frho-alpha}), we obtain that 
%%%%%%%%%%%%%%%%%%%%
\begin{eqnarray}
\tan 2\theta _{12}  \approx 
  \frac{{2\sqrt 2 b_0 }}{2d_0 - a_0}\left| {\frac{{\cos \alpha }}{{\cos \rho }}} \right|,
\label{Eq:C1 tan(2theta12) for complex m1=m2}
\end{eqnarray}
%%%%%%%%%%%%%%%%%%%%
where $d_0$ can be always chosen to be positive.  
From the constraint of $b_0\approx 0$, Eq.(\ref{Eq:C1 tan(2theta12) for complex m1=m2}) 
gives $\sin^22\theta_{12}= {\mathcal O}(1)$ if $2d_0 - a_0\approx 0$, thus requiring $a_0>0$, 
or if $\cos\rho\approx 0$ for $2d_0-a_0\neq 0$. 
This result is consistent with Eq.(\ref{Eq:C1 scale of neutrino masses}).
 From this observation, we find the following conditions:
%%%%%%%%%%%%%%%%%%%%
\begin{enumerate}
\item $\cos\rho$ is not suppressed and has moderate values including $\cos\rho=1$ for $a_0\approx 2d_0( > 0)$.  
The texture is given by Eq.(\ref{Eq:C1 Inverted1}).
\item $\cos\rho$ should be suppressed and $a_0-2d_0$ is not suppressed.  
The texture is given by Eq.(\ref{Eq:C1 Inverted1-2}).
\end{enumerate}
%%%%%%%%%%%%%%%%%%%%
For $x\neq 0$, which is the case with $m_1\approx -m_2$, since both $a_0$ and $d_0$ are not suppressed, either 
${\rm Im} \left( {e^{ - i\rho } x} \right)\sin \rho\approx 0$ or 
${\rm Re} \left( {e^{ - i\rho } x} \right)\cos \rho\approx 0$
should be suppressed. We have to require that
%%%%%%%%%%%%%%%%%%%%
\begin{enumerate}
\item $a_0+ 2d_0\approx0$ if ${\rm Im} \left( {e^{ - i\rho } x} \right)\sin\rho\approx 0$.  
The texture is given by Eq.(\ref{Eq:C1 Inverted2}).
\item $a_0- 2d_0\approx0$ if ${\rm Re} \left( {e^{ - i\rho } x} \right)\cos\rho\approx 0$. 
The texture is given by Eq.(\ref{Eq:C1 Inverted2-q}).
\end{enumerate}
%%%%%%%%%%%%%%%%%%%%

For the category (C2), the case of $x \approx 0$ can only be satisfied 
because $\vert x \vert \propto \varepsilon$ as a result of the approximate $\mu$-$\tau$ symmetry and 
Eq.(\ref{Eq:C1 scale of neutrino masses}) should be satisfied.  
Furthermore, to retain $\sin^2\theta_{13}\ll 1$, we have 
$\vert b_0\vert \ll\vert a_0\vert$ in $\tan 2\theta_{13}$ estimated in Eq.(\ref{Eq:App C2 mixing angles}),
which in turn further gives
%%%%%%%%%%%%%%%%%%%%
\begin{eqnarray}
\tan 2\theta _{12}  \approx 
  \frac{{X}}{\left(2\left|d_0\right| - \left|a_0\right|\right)\left(2\left|d_0\right| + \left|a_0\right|\right)}.
\label{Eq:C2 tan(2theta12) for complex m1=m2}
\end{eqnarray}
%%%%%%%%%%%%%%%%%%%% 
Since $\vert X\vert~={\mathcal O}(\varepsilon)$, we find that 
$\left| 2\left|d_0\right| - \left|a_0\right|\right| = {\mathcal O}\left( { \varepsilon } \right)$.
Therefore, we obtain conditions:
%%%%%%%%%%%%%%%%%%%%
\begin{enumerate}
\item $a_0- 2d_0\approx 0$ as well as $\cos\rho\neq 0$ from Eq.(\ref{Eq:C1 scale of neutrino masses}). The texture is given by Eq.(\ref{Eq:C2 Inverted}),
\item $a_0+ 2d_0\approx 0$ as well as $\sin\rho\neq 0$ from Eq.(\ref{Eq:C1 scale of neutrino masses}). The texture is given by Eq.(\ref{Eq:C2 Inverted-2}).
\end{enumerate}
%%%%%%%%%%%%%%%%%%%%

In both categories, the approximate $\mu$-$\tau$ symmetry assures that
%%%%%%%%%%%%%%%%%%%%
\begin{enumerate}
\item for the categories (C1) and (C2),  $\cos 2\theta_{23}$ is proportional to $\varepsilon$ and
the almost maximal atmospheric neutrino mixing naturally arises;
\item for the category (C1),  $\tan 2\theta_{13}$ is proportional to $\varepsilon$ and
the smallness of $\sin^2\theta_{13}$ naturally arises;
\item for the category (C2),  $x$ in $m_{1,2}$ is proportional to $\varepsilon$ and 
the smallness of $\Delta m^2_\odot(\equiv m^2_2-m^2_1)$ naturally arises because 
$\Delta m^2_\odot$ is proportional to $x$.  It is equivalent to refer to $X$ instead of $x$, which 
obviously gives the suppressed $\Delta m^2_\odot$ because of Eq.(\ref{Eq:App Mass from M^daggerM}).
\end{enumerate}
%%%%%%%%%%%%%%%%%%%%

In the next subsections, we estimate sizes of CP phases as functions of $\alpha$ and $\beta$
 valid up to ${\mathcal O}(\varepsilon)$. 
On the other hand, numerical analysis is based on our exact formulas shown in Sec.\ref{sec:3} 
without the perturbation of $\varepsilon$.  In each texture to be discussed, 
we use $m_0$ to denote the mass scale of neutrinos, $p,q$ satisfying $\vert p\vert= {\mathcal O}(1)$ 
and $\vert q\vert = {\mathcal O}(1)$ to respectively denote 
mass parameters for $M_{ee}$ and $M_{e\mu,e\tau}$ and $\eta$ to denote a tiny parameter, which provides 
$\Delta m^2_\odot/\vert\Delta m^2_{atm}\vert\ll 1$ for the category (C1) 
and $\sin^2\theta_{13}\ll 1$ for the category (C2). Roughly 
speaking, in the category (C1), $\Delta m^2_\odot/\vert\Delta m^2_{atm}\vert ={\mathcal O}(\eta^2)$ for the normal mass hierarchy 
and $={\mathcal O}(\eta)$ for the inverted mass hierarchy are satisfied.
The CP parameters that 
can be compared with those analyzed by experiments are $\delta_{CP}(=\delta+\rho)$ and $\phi_{1,2,3}$ 
used in $U_{PMNS}$.  We define 
the CP-violating Majorana phase to be $\phi_{CP}=\phi_3-\phi_2$ for the normal mass hierarchy and 
$\phi_{CP}=\phi_2-\phi_1$ for the inverted mass hierarchy. Estimated Dirac and Majorana phases are 
illustrated in Fugues as functions of $\sin^2\theta_{13}$.  Masses, mixing angles, and phases in each texture are 
estimated from Sec.\ref{subsec:4}.

%%%%%%%%%%%%%%%%%%%%%%%%%%%%%%%%%%%%%%%%%%%%%%%%%%%%%%%%%%%%%%%%
\subsection{\label{subsec:C1)}Category (C1)}
%%%%%%%%%%%%%%%%%%%%%%%%%%%%%%%%%%%%%%%%%%%%%%%%%%%%%%%%%%%%%%%%
%%%%%%%%%%%%%%%%%%%%%%%%%%%%%%%%%%%%%%%%%%%%%%%%%%%%%%%%%%%%%%%%
\subsubsection{\label{subsubsec:C1)N}Normal Mass Hierarchy}
%%%%%%%%%%%%%%%%%%%%%%%%%%%%%%%%%%%%%%%%%%%%%%%%%%%%%%%%%%%%%%%%
Our mass matrix $M_\nu$ can be parameterized by
%%%%%%%%%%%%%%%%%%%%
\begin{equation}
M^{(C1)N}_\nu = m_0\left( {\begin{array}{*{20}c}
   {p\eta } & {e^{i\alpha } \eta } & { - \sigma e^{i\alpha } \eta }  \\
   {e^{i\alpha } \eta } & 1 & {e_0}/{m_0}  \\
   { - \sigma e^{i\alpha } \eta } & {e_0}/{m_0} & 1  \\
\end{array}} \right) + \varepsilon\left( {\begin{array}{*{20}c}
   0 & {e^{i\beta } b^\prime_0 } & {\sigma e^{i\beta } b^\prime_0 }  \\
   {e^{i\beta } b^\prime_0 } & {d^\prime_0 } & 0  \\
   {\sigma e^{i\beta } b^\prime_0 } & 0 & { - d^\prime_0 }  \\
\end{array}} \right),
\label{Eq:C1 Normal}
\end{equation}
%%%%%%%%%%%%%%%%%%%%
where $\sigma e_0/m_0  = 1 - 2e^{2i\alpha}  \eta /p + {\mathcal O}(\varepsilon^2)$ 
from $s=1$ to give $m_1=0$ from Eq.(\ref{Eq:e from mu-tau symmetric det(Mnu) 2}).

\begin{figure}[!htbp]
\begin{flushleft}
%\includegraphics*[20mm,80mm][200mm,265mm]{phase.eps}
%\includegraphics*[20mm,90mm][200mm,265mm]{phase.eps}
%80==>90 title moves closer to figure
%\includegraphics*[20mm,90mm][200mm,265mm]{phase.eps}
%\includegraphics*[20mm,90mm][200mm,235mm]{phase.eps}
%265==>235 figure moves upward
\includegraphics*[30mm,193mm][195mm,255.2mm]{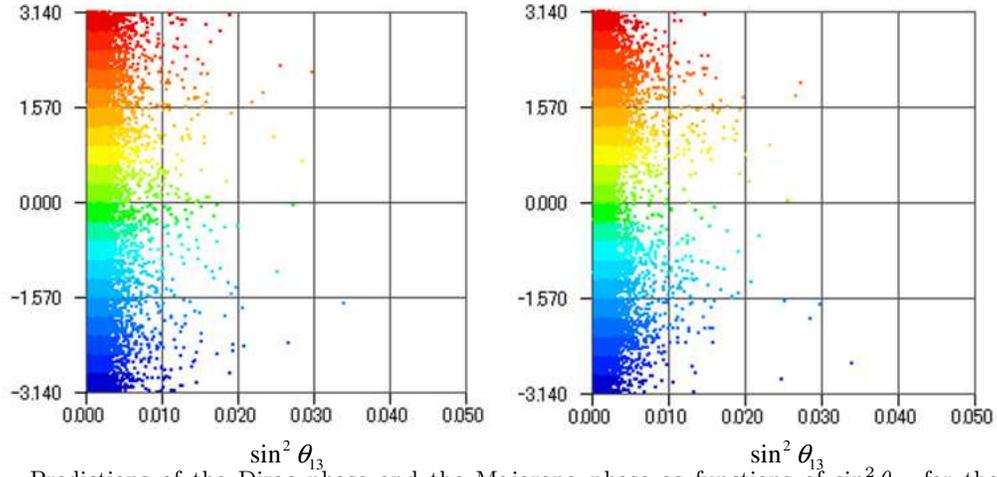}
\end{flushleft}
\vspace{-4mm}
\caption{Predictions of the Dirac phase and the Majorana phase as functions of $\sin^2\theta_{13}$
for the normal mass hierarchy given by $M^{(C1)N}_\nu$.}
\label{Fig:C1-N}
\end{figure}

We obtain that
%%%%%%%%%%%%%%%%%%%%
\begin{eqnarray}
\tan 2\theta _{12} e^{i\rho }  &\approx& \frac{{2\sqrt 2 e^{i\alpha } }}{{\frac{2}{p} - p}},
\nonumber\\
\tan 2\theta _{13} e^{ - i\delta }  &\approx& 
\frac{{\sqrt 2 \sigma {\varepsilon \left[ {b^\prime_0 
{\left[ {2{e^{ - i\beta }} + \eta \left( {p{e^{i\beta }} - 2\frac{1}{p}{e^{i\left( {2\alpha  - \beta } \right)}}} \right)} \right]}
 + \eta d^\prime_0 e^{ - i\alpha } } \right] }}}{{2m_0\left( {1 - 2\frac{\eta }{p}\cos 2\alpha } \right)}},
\label{Eq:C1 normal-theta12-theta13}
\end{eqnarray}
%%%%%%%%%%%%%%%%%%%%
where $\Delta$ and $\gamma$ are ${\mathcal O}(\varepsilon)$, from which
%%%%%%%%%%%%%%%%%%%%
\begin{equation}
\rho \approx \alpha,
\quad
\tan \delta  \approx
\frac{{{{b'}_0}\left[ {\left( {2 - p\eta } \right)\sin \beta  + \frac{{2\eta }}{p}\sin \left( {2\alpha  - \beta } \right)} \right] + \eta {{d'}_0}\sin \alpha }}{{{{b'}_0}\left[ {\left( {2 + p\eta } \right)\cos \beta  - \frac{{2\eta }}{p}\cos \left( {2\alpha  - \beta } \right)} \right] + \eta {{d'}_0}\cos \alpha }}
\label{Eq:C1 normal-delta-rho}
\end{equation}
%%%%%%%%%%%%%%%%%%%%
are derived. Since $\eta$ in $\delta$ is phenomenologically suppressed, we find that
%%%%%%%%%%%%%%%%%%%%
\begin{equation}
\rho \approx \alpha,
\quad
\delta \approx \beta.
\label{Eq:C1 normal-delta-rho-approx}
\end{equation}
%%%%%%%%%%%%%%%%%%%%
It should be noted that Dirac CP violation is controlled by $\delta_{CP}\approx \alpha+\beta$ while Majorana CP violation 
is associated with neutrino masses, which are given by
%%%%%%%%%%%%%%%%%%%%
\begin{equation}
 m_2 e^{ - 2i\phi _2 }  \approx \left( {p + \frac{2}{p}} \right)e^{2i\rho } \eta m_0,
\quad
 m_3 e^{ - 2i\phi _3 }  \approx 2\left( {1 - \frac{\eta }{p}e^{2i\alpha } } \right)m_0,
\label{Eq:C1 normal-phi2-phi3}
\end{equation}
%%%%%%%%%%%%%%%%%%%%
leading to
%%%%%%%%%%%%%%%%%%%%
\begin{equation}
\phi _2  \approx  - \rho,
\quad
\phi _3  \approx 0.
\label{Eq:C1 normal-MajoranaPhase}
\end{equation}
%%%%%%%%%%%%%%%%%%%%
Therefore, Majorana CP violation is controlled by $\phi_{CP}=\phi_3-\phi_2$:
%%%%%%%%%%%%%%%%%%%%
\begin{equation}
\phi_{CP} (\approx \rho)\approx \alpha,
\label{Eq:C1 normal-CPMajoranaPhase}
\end{equation}
%%%%%%%%%%%%%%%%%%%%
which is shown in FIG.\ref{Fig:C1-N} as function of $\sin^2\theta_{13}$, where no constraint on the size of 
$\phi_{CP}$ is found.  In other words, the maximal CP violation signaled by $\phi_{CP}\approx \pi/2$ is allowed.

%%%%%%%%%%%%%%%%%%%%%%%%%%%%%%%%%%%%%%%%%%%%%%%%%%%%%%%%%%%%%%%%
\subsubsection{\label{subsubsec:C1)I-A}Inverted mass hierarchy I A ($m_1\approx m_2$)}
%%%%%%%%%%%%%%%%%%%%%%%%%%%%%%%%%%%%%%%%%%%%%%%%%%%%%%%%%%%%%%%%
Our mass matrix $M_\nu$ can be parameterized by
%%%%%%%%%%%%%%%%%%%%
\begin{equation}
M^{(C1)IA}_\nu   = m_0\left( {\begin{array}{*{20}c}
   {2 - p\eta } & {e^{i\alpha } \eta } & { - \sigma e^{i\alpha } \eta }  \\
   {e^{i\alpha } \eta } & 1 & {e_0}/{m_0}  \\
   { - \sigma e^{i\alpha } \eta } & {e_0}/{m_0} & 1  \\
\end{array}} \right) + \varepsilon\left( {\begin{array}{*{20}c}
   0 & {e^{i\beta } b^\prime_0 } & {\sigma e^{i\beta } b^\prime_0 }  \\
   {e^{i\beta } b^\prime_0 } & {d^\prime_0 } & 0  \\
   {\sigma e^{i\beta } b^\prime_0 } & 0 & { - d^\prime_0 }  \\
\end{array}} \right),
\label{Eq:C1 Inverted1}
\end{equation}
%%%%%%%%%%%%%%%%%%%%
where $\sigma e_0/m_0  = -1 + {\mathcal O}(\varepsilon^2)$ from $s=-1$ 
to give $m_3=0$ from Eq.(\ref{Eq:e from mu-tau symmetric det(Mnu) 1}).

\begin{figure}[!htbp]
\begin{flushleft}
%\includegraphics*[20mm,80mm][200mm,265mm]{phase.eps}
%\includegraphics*[20mm,90mm][200mm,265mm]{phase.eps}
%80==>90 title moves closer to figure
%\includegraphics*[20mm,90mm][200mm,265mm]{phase.eps}
%\includegraphics*[20mm,90mm][200mm,235mm]{phase.eps}
%265==>235 figure moves upward
\includegraphics*[30mm,193mm][195mm,255.2mm]{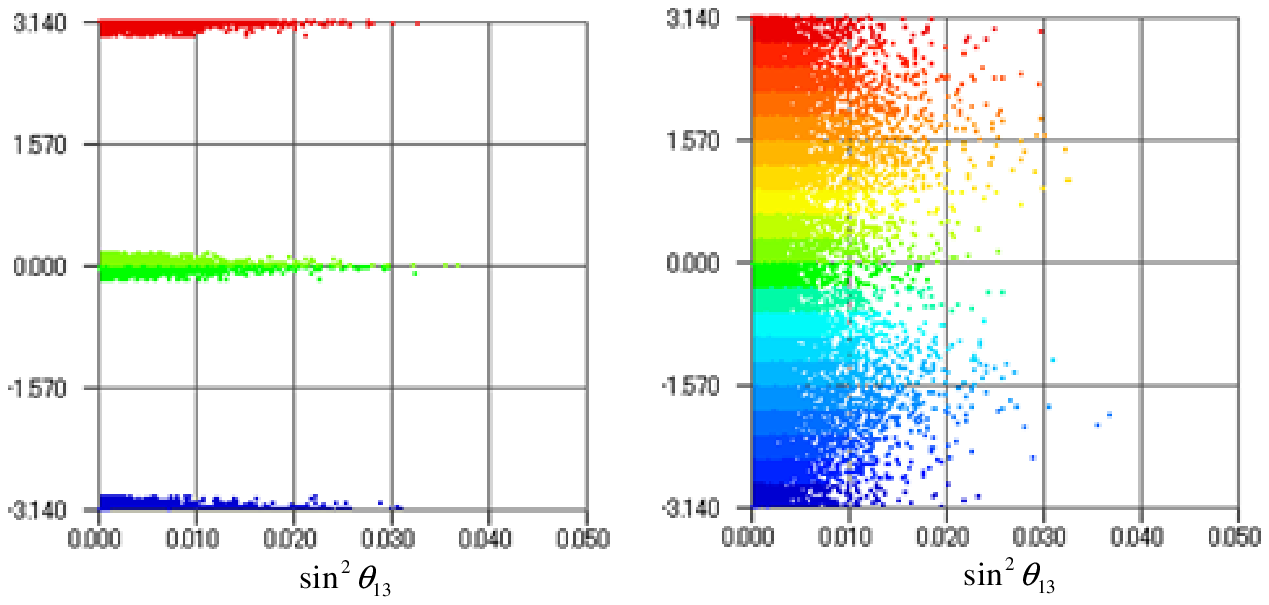}
\end{flushleft}
\vspace{-4mm}
\caption{The same as in FIG.\ref{Fig:C1-N} but 
for the inverted mass hierarchy with $m_1\approx m_2$ given by $M^{(C1)IA}_\nu$.}
\label{Fig:C1-IA}
\end{figure}

We obtain that
%%%%%%%%%%%%%%%%%%%%
\begin{eqnarray}
\tan 2\theta _{12} e^{i\rho }  &\approx& 
2\sqrt 2 \left( {\frac{\cos \alpha}{p}  - i\frac{\eta \sin \alpha}{{4 - p\eta }} } \right),
\nonumber\\
\tan 2\theta _{13} e^{ - i\delta }  &\approx& 
 - \frac{\sqrt 2 \sigma \left[ {\varepsilon\left(2-p\eta\right) b^\prime_0 e^{i\beta }  + \eta \left( {\varepsilon d^\prime_0 e^{ - i\alpha }  - 4m_0\left( {\Delta  - i\gamma } \right)\cos \alpha } \right)} \right]}{2m_0\left(1 - p\eta\right)}.
\nonumber\\
\label{Eq:C1 inverted1-theta12-theta13}
\end{eqnarray}
%%%%%%%%%%%%%%%%%%%%
Owing to the phenomenological requirement of 
$\sin^22\theta_{12}={\mathcal O}(1)$, $\cos\alpha ={\mathcal O}(1)$ should be realized.
It is found that
%%%%%%%%%%%%%%%%%%%%
\begin{equation}
\tan \rho  \approx  - \frac{p\eta }{4 - p\eta }\tan \alpha,
\quad
\tan \delta  \approx  - \frac{\varepsilon\left(2-p\eta\right) b^\prime_0 \sin \beta  - \eta \left( {\varepsilon d^\prime_0 \sin \alpha  - 4m_0\gamma \cos \alpha } \right)}{\varepsilon\left(2-p\eta\right) b^\prime_0 \cos \beta  + \eta \left( {\varepsilon d^\prime_0  - 4m_0\Delta } \right)\cos \alpha }.
\label{Eq:C1 inverted1-delta-rho}
\end{equation}
%%%%%%%%%%%%%%%%%%%%
Since $\cos\alpha={\mathcal O}(1)$ and the terms proportional to $\eta$ in $\tan \delta$ can be neglected, 
we observe that 
%%%%%%%%%%%%%%%%%%%%
\begin{equation}
\rho\approx 0, 
\quad
\delta\approx -\beta,
\label{Eq:C1 inverted1-delta-rho-approx}
\end{equation}
%%%%%%%%%%%%%%%%%%%%
leading to
%%%%%%%%%%%%%%%%%%%%
\begin{equation}
\delta_{CP} \approx -\beta.
\label{Eq:C1 inverted1-delta-CP-approx}
\end{equation}
%%%%%%%%%%%%%%%%%%%%
Majorana CP violation is associated with neutrino masses, which are calculated to be:
%%%%%%%%%%%%%%%%%%%%
\begin{eqnarray}
 m_1 e^{ - 2i\phi _1 }  \approx 
e^{i\rho } \left( {\frac{4\cos \rho  -\eta e^{i\rho } p}{2}  - \frac{\sqrt 2 \eta e^{i\alpha }}{{\sin 2\theta _{12} }}} \right)m_0 ,
\nonumber\\
 m_2 e^{ - 2i\phi _2 }  \approx 
e^{i\rho } \left( {\frac{4\cos \rho  -\eta e^{i\rho } p }{2}  + \frac{\sqrt 2 \eta e^{i\alpha }}{{\sin 2\theta _{12} }}} \right)m_0 ,
\label{Eq:C1 inverted1-phi2-phi3}
\end{eqnarray}
%%%%%%%%%%%%%%%%%%%%
from which
%%%%%%%%%%%%%%%%%%%%
\begin{equation}
\phi _1  \approx \phi _2  \approx 0,
\label{Eq:C1 inverted1-MajoranaPhase}
\end{equation}
%%%%%%%%%%%%%%%%%%%%
because of $\rho\approx 0$.
Therefore, Majorana CP violation is controlled by $\phi_{CP}=\phi_2-\phi_1$:
%%%%%%%%%%%%%%%%%%%%
\begin{equation}
\phi_{CP} \approx 0,
\label{Eq:C1 inverted1-CPMajoranaPhase}
\end{equation}
%%%%%%%%%%%%%%%%%%%%
which gives FIG.\ref{Fig:C1-IA}.

%%%%%%%%%%%%%%%%%%%%%%%%%%%%%%%%%%%%%%%%%%%%%%%%%%%%%%%%%%%%%%%%
\subsubsection{\label{subsubsec:C1)I-B}Inverted mass hierarchy I B ($m_1\approx m_2$)}
%%%%%%%%%%%%%%%%%%%%%%%%%%%%%%%%%%%%%%%%%%%%%%%%%%%%%%%%%%%%%%%%
Our mass matrix $M_\nu$ can be parameterized by
%%%%%%%%%%%%%%%%%%%%
\begin{equation}
M^{(C1)IB}_\nu   = m_0\left( {\begin{array}{*{20}c}
   {-2 + p\eta } & {e^{i\alpha } \eta } & { - \sigma e^{i\alpha } \eta }  \\
   {e^{i\alpha } \eta } & 1 & {e_0}/{m_0}  \\
   { - \sigma e^{i\alpha } \eta } & {e_0}/{m_0} & 1  \\
\end{array}} \right) + \varepsilon\left( {\begin{array}{*{20}c}
   0 & {e^{i\beta } b^\prime_0 } & {\sigma e^{i\beta } b^\prime_0 }  \\
   {e^{i\beta } b^\prime_0 } & {d^\prime_0 } & 0  \\
   {\sigma e^{i\beta } b^\prime_0 } & 0 & { - d^\prime_0 }  \\
\end{array}} \right),
\label{Eq:C1 Inverted1-2}
\end{equation}
%%%%%%%%%%%%%%%%%%%%
where $\sigma e_0/m_0  = -1 + {\mathcal O}(\varepsilon^2)$ from $s=1$ 
to give $m_3=0$ from Eq.(\ref{Eq:e from mu-tau symmetric det(Mnu) 1}).
The sign of $a_0$ differs from the one for Eq.(\ref{Eq:C1 Inverted1}). This sign difference 
converts $a_0  + 2d_0$ into $-a_0  + 2d_0$ and in turn exchanges the role of $\cos\rho$ and $\sin\rho$ in 
Eq.(\ref{Eq:C1 scale of neutrino masses});
%%%%%%%%%%%%%%%%%%%%

\begin{figure}[!htbp]
\begin{flushleft}
%\includegraphics*[20mm,80mm][200mm,265mm]{phase.eps}
%\includegraphics*[20mm,90mm][200mm,265mm]{phase.eps}
%80==>90 title moves closer to figure
%\includegraphics*[20mm,90mm][200mm,265mm]{phase.eps}
%\includegraphics*[20mm,90mm][200mm,235mm]{phase.eps}
%265==>235 figure moves upward
\includegraphics*[30mm,193mm][195mm,255.2mm]{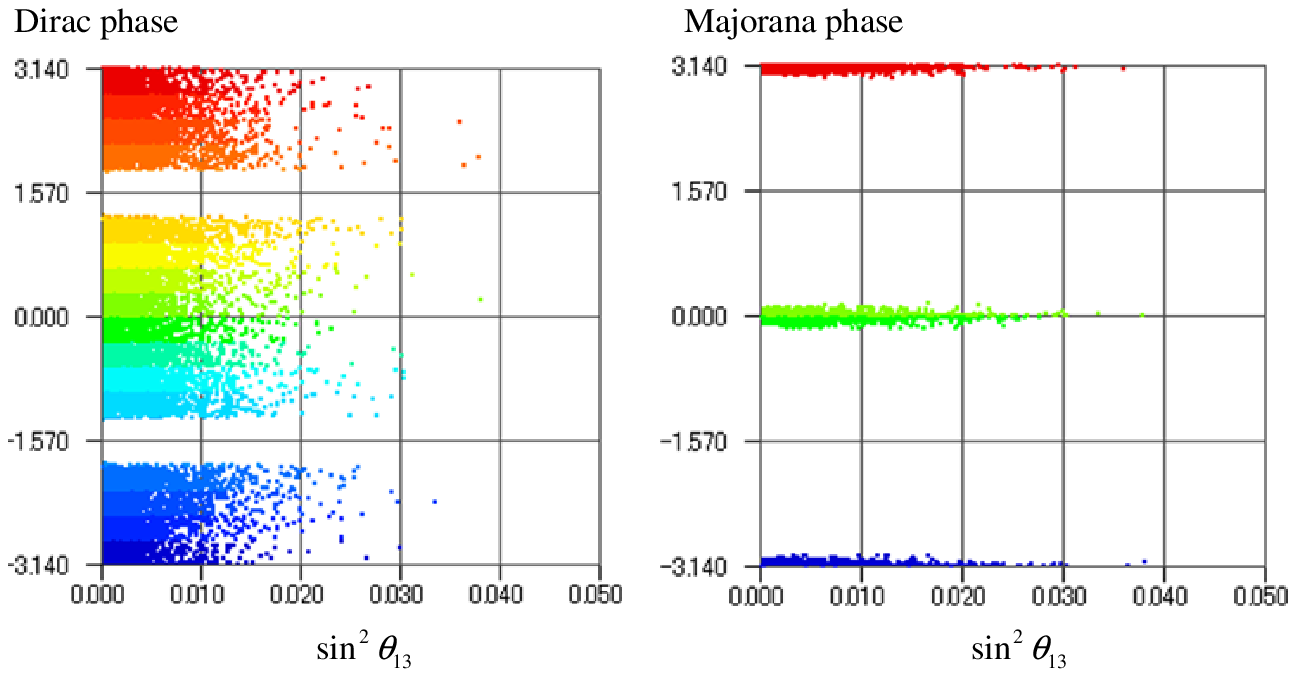}
\end{flushleft}
\vspace{-4mm}
\caption{The same as in FIG.\ref{Fig:C1-N} but
for the inverted mass hierarchy with $m_1\approx m_2$ given by $M^{(C1)IB}_\nu$.}
\label{Fig:C1-IB}
\end{figure}

We obtain that
%%%%%%%%%%%%%%%%%%%%
\begin{eqnarray}
\tan 2\theta _{12} e^{i\rho }  &\approx& 
2\sqrt 2 \left( { - i\frac{\sin \alpha}{p}  + \frac{{\eta \cos \alpha}}{{4 - p\eta }} } \right),
\nonumber\\
\tan 2\theta _{13} e^{ - i\delta }  &\approx& 
-\sqrt 2 \sigma \frac{{\varepsilon \left( { - \left(2-p\eta\right)b^\prime_0 e^{i\beta }  + \eta d^\prime_0 e^{ - i\alpha } } \right) 
+ 4m_0i\left( \Delta  - i\gamma \right)\eta\sin \alpha}}{{2m_0\left(1 - p\eta\right) }}.
\nonumber\\
\label{C1 Inverted1-2-theta12-theta13}
\end{eqnarray}
%%%%%%%%%%%%%%%%%%%%
Owing to the phenomenological requirement of 
$\sin^22\theta_{12}={\mathcal O}(1)$, $\vert\sin\alpha\vert ={\mathcal O}(1)$ should be realized.
It is found that
%%%%%%%%%%%%%%%%%%%%
\begin{eqnarray}
\tan \rho  &\approx&  
 - \frac{{4 - p\eta }}{{p\eta }}\tan \alpha, 
\nonumber\\
\tan \delta  &\approx&  
 - \frac{\varepsilon\left(2-p\eta\right) b'_0 \sin \beta  + \eta \left( {\varepsilon d'_0  - 4m_0 \Delta } \right)\sin \alpha }{\varepsilon\left(2-p\eta\right) b'_0 \cos \beta  - \eta \left( {\varepsilon d'_0 \cos \alpha  + 4m_0 \gamma \sin \alpha } \right)}.
\label{C1 Inverted1-2-delta-rho}
\end{eqnarray}
%%%%%%%%%%%%%%%%%%%%
Since $\vert\sin\alpha\vert={\mathcal O}(1)$ and $\eta$ is phenomenologically suppressed, we observe that 
%%%%%%%%%%%%%%%%%%%%
\begin{equation}
\rho\approx \pm \pi/2, 
\quad
\delta\approx -\beta,
\label{Eq:C1 inverted1-2-delta-rho-approx}
\end{equation}
%%%%%%%%%%%%%%%%%%%%
leading to
%%%%%%%%%%%%%%%%%%%%
\begin{equation}
\delta_{CP} \approx -\beta\pm\frac{\pi}{2}.
\label{Eq:C1 inverted1-2-delta-CP-approx}
\end{equation}
%%%%%%%%%%%%%%%%%%%%
Majorana CP violation is associated with neutrino masses, which are calculated to be:
%%%%%%%%%%%%%%%%%%%%
\begin{eqnarray}
 m_1 e^{ - 2i\phi _1 }  &\approx& 
 - e^{i\rho } \left( {2i\sin \rho  - \frac{{p\eta e^{i\rho}}}{2} + \frac{{\sqrt 2 e^{i\alpha } }}{{\sin 2\theta _{12} }}\eta } \right)m_0,
\nonumber\\
 m_2 e^{ - 2i\phi _2 }  &\approx& 
 - e^{i\rho } \left( {2i\sin \rho  - \frac{{p\eta e^{i\rho}}}{2} - \frac{{\sqrt 2 e^{i\alpha } }}{{\sin 2\theta _{12} }}\eta } \right)m_0, 
\label{C1 Inverted1-2-phi2-phi3}
\end{eqnarray}
%%%%%%%%%%%%%%%%%%%%
from which
%%%%%%%%%%%%%%%%%%%%
\begin{equation}
\phi _1  \approx \phi _2  \approx 0,
\label{C1 Inverted1-2-MajoranaPhase}
\end{equation}
%%%%%%%%%%%%%%%%%%%%
because of Eq.(\ref{Eq:C1 inverted1-2-delta-rho-approx}). 
Therefore, Majorana CP violation is controlled by $\phi_{CP}=\phi_2-\phi_1$:
%%%%%%%%%%%%%%%%%%%%
\begin{equation}
\phi_{CP} \approx 0.
\label{C1 Inverted1-2-CPMajoranaPhase}
\end{equation}
%%%%%%%%%%%%%%%%%%%%
which gives FIG.\ref{Fig:C1-IB}.

%%%%%%%%%%%%%%%%%%%%%%%%%%%%%%%%%%%%%%%%%%%%%%%%%%%%%%%%%%%%%%%%
\subsubsection{\label{subsubsec:C1 II-A}Inverted mass hierarchy I\hspace{-.1em}I A ($m_1\approx -m_2$)}
%%%%%%%%%%%%%%%%%%%%%%%%%%%%%%%%%%%%%%%%%%%%%%%%%%%%%%%%%%%%%%%%
Our mass matrix $M_\nu$ can be parameterized by
%%%%%%%%%%%%%%%%%%%%
\begin{equation}
M^{(C1)I\hspace{-.1em}IA}_\nu  = m_0 \left( {\begin{array}{*{20}c}
   { - 2 + \eta } & {e^{i\alpha } q} & { - \sigma e^{i\alpha } q}  \\
   {e^{i\alpha } q} & 1 & {e_0}/{m_0}  \\
   { - \sigma e^{i\alpha } q} & {e_0}/{m_0} & 1  \\
\end{array}} \right) + \varepsilon\left( {\begin{array}{*{20}c}
   0 & {e^{i\beta } b^\prime_0 } & {\sigma e^{i\beta } b^\prime_0 }  \\
   {e^{i\beta } b^\prime_0 } & {d^\prime_0 } & 0  \\
   {\sigma e^{i\beta } b^\prime_0 } & 0 & { - d^\prime_0 }  \\
\end{array}} \right),
\label{Eq:C1 Inverted2}
\end{equation}
%%%%%%%%%%%%%%%%%%%%
where $\sigma e_0/m_0  = -1 + {\mathcal O}(\varepsilon^2)$ from $s=1$ 
to give $m_3=0$ from Eq.(\ref{Eq:e from mu-tau symmetric det(Mnu) 1}).

\begin{figure}[!htbp]
\begin{flushleft}
%\includegraphics*[20mm,80mm][200mm,265mm]{phase.eps}
%\includegraphics*[20mm,90mm][200mm,265mm]{phase.eps}
%80==>90 title moves closer to figure
%\includegraphics*[20mm,90mm][200mm,265mm]{phase.eps}
%\includegraphics*[20mm,90mm][200mm,235mm]{phase.eps}
%265==>235 figure moves upward
\includegraphics*[30mm,193mm][195mm,255.2mm]{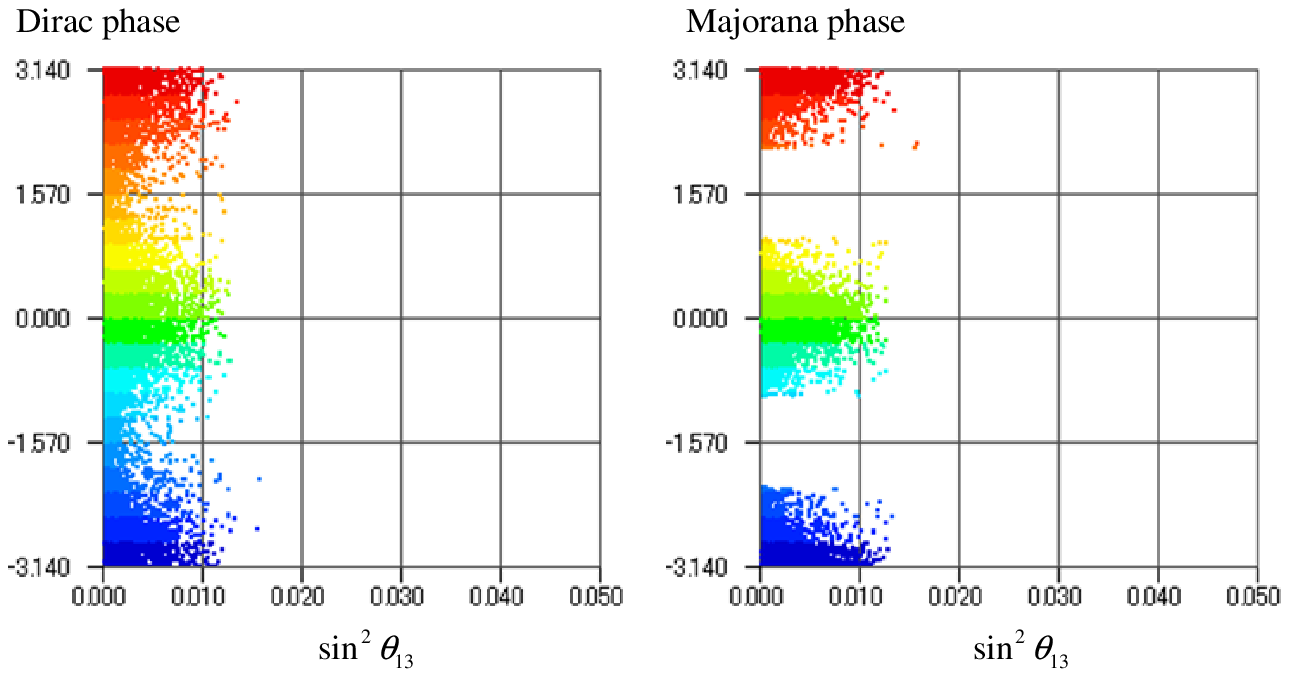}
\end{flushleft}
\vspace{-4mm}
\caption{The same as in FIG.\ref{Fig:C1-N} but 
for the inverted mass hierarchy with $m_1\approx -m_2$ given by $M^{(C1)I\hspace{-.1em}IA}_\nu$.}
\label{Fig:C1-IIA}
\end{figure}

We obtain  that
%%%%%%%%%%%%%%%%%%%%
\begin{eqnarray}
&&\tan 2\theta _{12} e^{i\rho }  \approx 
2\sqrt 2 q\left( {\frac{{\cos \alpha }}{{4 - \eta }} - i\frac{{\sin \alpha }}{\eta }} \right),
\nonumber\\
&&\tan 2\theta _{13} e^{ - i\delta }  \approx 
\nonumber\\
&& ~\frac{
\sqrt 2 \sigma \left[ 
\varepsilon \left( {\left(2-\eta\right)b^\prime_0 e^{i\beta }  - qd^\prime_0 e^{ - i\alpha } } \right)
 - m_0qi\left(\Delta-i\gamma\right)\left( \left(4-\eta\right)\sin\alpha + i\eta\cos\alpha \right)
 \right]
  }
 {
 m_0\left[{2\left( {1 - \eta } \right) + q^2 }\right]
 }.
\nonumber\\ 
\label{Eq:C1 inverted2-theta12-theta13}
\end{eqnarray}
%%%%%%%%%%%%%%%%%%%%
Owing to the phenomenological requirement of 
$\sin^22\theta_{12}={\mathcal O}(1)$, $\vert \sin \alpha \vert \leq {\mathcal O}(\eta)$ should be satisfied.
It is found that
%%%%%%%%%%%%%%%%%%%%
\begin{eqnarray}
&&\tan \rho  \approx
\frac{{\eta  - 4}}{\eta }\tan \alpha, 
\nonumber\\
&&\tan \delta  \approx
\nonumber\\
&& ~ - \frac{\varepsilon \left( {\left(2-\eta\right)b'_0 \sin \beta  + qd'_0 \sin \alpha } \right) - m_0 q\left( {4\Delta \sin \alpha  + \gamma \eta \cos \alpha } \right)}{\varepsilon \left( {\left(2-\eta\right)b'_0 \cos \beta  - qd'_0 \cos \alpha } \right) + m_0 q\left( {\eta \Delta \cos \alpha  - 4\gamma \sin \alpha } \right)}.
\label{Eq:C1 inverted2-delta-rho}
\end{eqnarray}
%%%%%%%%%%%%%%%%%%%%
Majorana CP violation is associated with neutrino masses, which are calculated to be:
%%%%%%%%%%%%%%%%%%%%
\begin{eqnarray}
 m_1 e^{ - 2i\phi _1 }  &\approx& 
-e^{i\rho } \left( {\frac{{\sqrt 2 e^{i\alpha } q}}{{\sin 2\theta _{12} }}+ \frac{{4i\sin \rho }-e^{i\rho } \eta}{2}} \right)m_0,
\nonumber\\
 m_2 e^{ - 2i\phi _2 }  &\approx& 
 e^{i\rho } \left( {\frac{{\sqrt 2 e^{i\alpha } q}}{{\sin 2\theta _{12} }}-\frac{{4i\sin \rho }-e^{i\rho } \eta }{2}} \right)m_0.
\label{C1 Inverted12-phi2-phi3}
\end{eqnarray}
%%%%%%%%%%%%%%%%%%%%
To satisfy $\Delta m^2_\odot/\vert\Delta m^2_{atm}\vert\ll 1$, 
we require that $\vert \sin\rho\sin\alpha\vert\leq {\mathcal O}( \eta )$. Since $\vert \sin\alpha\vert \leq {\mathcal O}(\eta) $ , 
$\vert\sin\rho\vert$ can be large enough to affect the size of CP violation as far as $m_1\approx -m_2$ is kept. The difference of $\phi_1$ and $\phi_2$ is 
enhanced for smaller $q$ and larger $\sin\rho$.
Therefore, this texture provides larger effect of Majorana CP violation controlled by $\phi_{CP}=\phi_2-\phi_1$ as shown 
in FIG.\ref{Fig:C1-IIA}.

%%%%%%%%%%%%%%%%%%%%%%%%%%%%%%%%%%%%%%%%%%%%%%%%%%%%%%%%%%%%%%%%
\subsubsection{\label{subsubsec:C1)II-B}Inverted mass hierarchy I\hspace{-.1em}I B ($m_1\approx -m_2$)}
%%%%%%%%%%%%%%%%%%%%%%%%%%%%%%%%%%%%%%%%%%%%%%%%%%%%%%%%%%%%%%%%
Our mass matrix $M_\nu$ can be parameterized by
%%%%%%%%%%%%%%%%%%%%
\begin{equation}
M^{(C1)I\hspace{-.1em}IB}_\nu  = m_0 \left( {\begin{array}{*{20}c}
   { 2 - \eta } & {e^{i\alpha } q} & { - \sigma e^{i\alpha } q}  \\
   {e^{i\alpha } q} & 1 & {e_0}/{m_0}  \\
   { - \sigma e^{i\alpha } q} & {e_0}/{m_0} & 1  \\
\end{array}} \right) + \varepsilon\left( {\begin{array}{*{20}c}
   0 & {e^{i\beta } b^\prime_0 } & {\sigma e^{i\beta } b^\prime_0 }  \\
   {e^{i\beta } b^\prime_0 } & {d^\prime_0 } & 0  \\
   {\sigma e^{i\beta } b^\prime_0 } & 0 & { - d^\prime_0 }  \\
\end{array}} \right),
\label{Eq:C1 Inverted2-q}
\end{equation}
%%%%%%%%%%%%%%%%%%%%
where $\sigma e_0/m_0  = -1 + {\mathcal O}(\varepsilon^2)$ from $s=-1$ 
to give $m_3=0$ from Eq.(\ref{Eq:e from mu-tau symmetric det(Mnu) 1}).
Similarly to the relation between 
Eq.(\ref{Eq:C1 Inverted1}) and  Eq.(\ref{Eq:C1 Inverted1-2}), 
the sign of $a_0$ differs from the one for Eq.(\ref{Eq:C1 Inverted2}).

\begin{figure}[!htbp]
\begin{flushleft}
%\includegraphics*[20mm,80mm][200mm,265mm]{phase.eps}
%\includegraphics*[20mm,90mm][200mm,265mm]{phase.eps}
%80==>90 title moves closer to figure
%\includegraphics*[20mm,90mm][200mm,265mm]{phase.eps}
%\includegraphics*[20mm,90mm][200mm,235mm]{phase.eps}
%265==>235 figure moves upward
\includegraphics*[30mm,193mm][195mm,255.2mm]{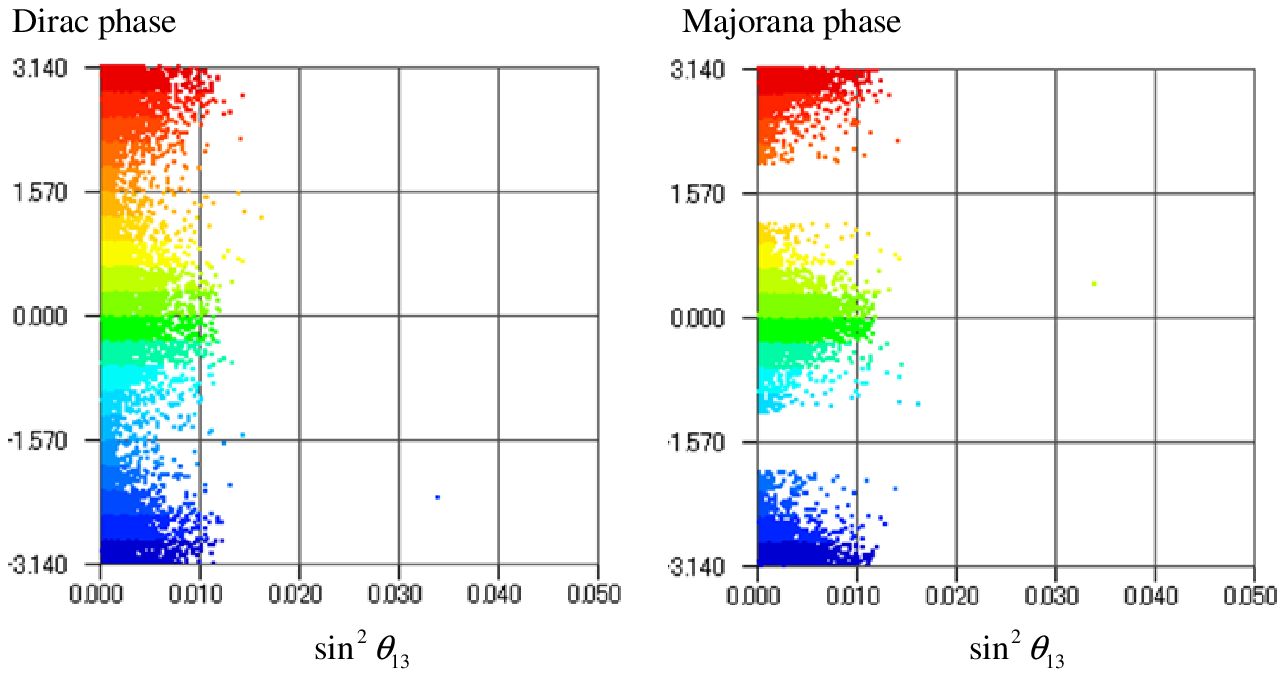}
\end{flushleft}
\vspace{-4mm}
\caption{The same as in FIG.\ref{Fig:C1-N} but 
for the inverted mass hierarchy with $m_1\approx -m_2$ given by $M^{(C1)I\hspace{-.1em}IB}_\nu$.}
\label{Fig:C1-IIB}
\end{figure}

The predicted results are very similar to those for the previous texture.
We obtain  that
%%%%%%%%%%%%%%%%%%%%
\begin{eqnarray}
&&\tan 2\theta _{12} e^{i\rho }  \approx 
2\sqrt 2 q\left( {\frac{{\cos \alpha }}{\eta } - i\frac{{ \sin \alpha }}{{4 - \eta }}} \right),
\nonumber\\
&&\tan 2\theta _{13} e^{ - i\delta }  \approx
\nonumber\\
&&~- \frac{
\sqrt 2 \sigma \left[ {\varepsilon \left( {\left(2-\eta\right)e^{i\beta } b'_0  
 + e^{ - i\alpha } qd'_0 } \right) - m_0q\left( {\Delta  - i\gamma } \right)
 \left( {\left(4-\eta\right)\cos \alpha  - i\eta \sin \alpha } \right)} \right]
 }
 {{m_0\left[\left( {2 - \eta } \right)^2  + 2q^2 \right]}}.
\nonumber\\
\label{Eq:C1 inverted2-2-theta12-theta13}
\end{eqnarray}
%%%%%%%%%%%%%%%%%%%%
Owing to the phenomenological constraint of 
$\sin^22\theta_{12}={\mathcal O}(1)$, $\vert \cos\alpha\vert \leq {\mathcal O}(\eta) $ 
should be satisfied and 
is numerically signaled by $\alpha\approx \pm \pi/2$ found 
in FIG.\ref{Fig:C1-IIB}.
It is found that
%%%%%%%%%%%%%%%%%%%%
\begin{eqnarray}
\tan \rho  &\approx&
 - \frac{\eta }{{4 - \eta }}\tan \alpha.
\nonumber\\
\tan \delta  &\approx&  
 - \frac{{\varepsilon \left( {\left(2-\eta\right)b'_0 \sin \beta  - qd'_0 \sin \alpha } \right) 
 + m_0 q\left( {\left(4-\eta\right)\gamma \cos \alpha  + \eta \Delta \sin \alpha } \right)}}
 {{\varepsilon \left( {\left(2-\eta\right)b'_0 \cos \beta  + qd'_0 \cos \alpha } \right) - m_0 q\left( {\left(4-\eta\right)\Delta \cos \alpha  - \gamma \eta \sin \alpha } \right)}}.
\nonumber\\
\label{Eq:C1 inverted2-2-delta-rho}
\end{eqnarray}
%%%%%%%%%%%%%%%%%%%%
Majorana CP violation is associated with neutrino masses, which are calculated to be:
%%%%%%%%%%%%%%%%%%%%
\begin{eqnarray}
 m_1 e^{ - 2i\phi _1 }  &\approx& 
-e^{i\rho } \left( {\frac{\sqrt 2 qe^{i\alpha }}{{\sin 2\theta _{12} }} 
- \frac{4\cos \rho  - \eta e^{i\rho }}{2}} \right)m_0,
\nonumber\\
 m_2 e^{ - 2i\phi _2 }  &\approx& 
e^{i\rho } \left( { \frac{\sqrt 2 qe^{i\alpha } }{{\sin 2\theta _{12} }} 
+ \frac{4\cos \rho  - \eta e^{i\rho }}{2}} \right)m_0.
\label{C1 Inverted12-2-phi2-phi3}
\end{eqnarray}
%%%%%%%%%%%%%%%%%%%%
To satisfy $\Delta m^2_\odot/\vert\Delta m^2_{atm}\vert\ll 1$, 
we require that $\vert \cos\rho\cos\alpha\vert\leq {\mathcal O}( \eta )$. Since $\vert \cos\alpha\vert \leq {\mathcal O}(\eta) $, 
$\vert \cos\rho\vert$ can also be large. As far as $\vert \cos\rho\vert={\mathcal O}(1)$ is maintained, 
the difference of $\phi_1$ and $\phi_2$ is 
enhanced for smaller $q$ and larger $\cos\rho$. The size of $\phi_{CP}$ is shown in 
FIG.\ref{Fig:C1-IIB}, which indicates larger effect of Majorana CP violation controlled by $\phi_{CP}=\phi_2-\phi_1$.  

%%%%%%%%%%%%%%%%%%%%%%%%%%%%%%%%%%%%%%%%%%%%%%%%%%%%%%%%%%%%%%%%
\subsection{\label{subsec:C2)}Category (C2)}
%%%%%%%%%%%%%%%%%%%%%%%%%%%%%%%%%%%%%%%%%%%%%%%%%%%%%%%%%%%%%%%%
There are only textures giving the inverted mass hierarchy with $m_1\approx m_2$
because of the generic smallness of $x$ in Eq.(\ref{Eq:App C2 inverted masses and Majorana phases}).
To ensure another smallness of $\sin\theta_{13}$ requires $b_0\approx 0$ 
 as indicated by Eqs.(\ref{Eq:App MixingAngle from M}) and (\ref{Eq:App masses and mixing angles M}).  
 The inverted mass hierarchy with 
$m_1\approx -m_2$, corresponding to Eq.(\ref{Eq:C1 Inverted2}) cannot be accepted because of
$\vert b_0/m_0\vert=\vert q\vert={\mathcal O}(1)$.
A possible texture giving the normal mass hierarchy, namely, corresponding to 
Eq.(\ref{Eq:C1 Normal}) cannot be accepted.  This is because the condition on $m_1$=0 for the category (C2) is given by 
Eq.(\ref{Eq:e from mu-tau symmetric det(Mnu) 2}) leading to 
$\sigma e_0  = d_0 - 2b^2_0 /a_0 + {\mathcal O}(\varepsilon^2)$, which, however, 
gives $\tan 2\theta_{12}\approx 0$ in Eq.(\ref{Eq:App C2 mixing angles}) 
because its denominator is not suppressed owing to $a_0(=p\eta)\approx 0$ and $b_0\approx 0$. 
%%%%%%%%%%%%%%%%%%%%%%%%%%%%%%%%%%%%%%%%%%%%%%%%%%%%%%%%%%%%%%%%
\subsubsection{\label{subsubsec:C2)I-A}Inverted mass hierarchy I A ($m_1\approx m_2$)}
%%%%%%%%%%%%%%%%%%%%%%%%%%%%%%%%%%%%%%%%%%%%%%%%%%%%%%%%%%%%%%%%
Our mass matrix $M_\nu$ can be parameterized by
%%%%%%%%%%%%%%%%%%%%
\begin{equation}
M^{(C2)IA}_\nu = m_0\left( {\begin{array}{*{20}c}
   {2 - p\eta } & {e^{i\alpha } \eta } & { - \sigma e^{i\alpha } \eta }  \\
   {e^{i\alpha } \eta } & 1 & {e_0}/{m_0}  \\
   { - \sigma e^{i\alpha } \eta } & {e_0}/{m_0} & 1  \\
\end{array}} \right) + \varepsilon\left( {\begin{array}{*{20}c}
   0 & {e^{i\beta } b^\prime_0 } & {\sigma e^{i\beta } b^\prime_0 }  \\
   {e^{i\beta } b^\prime_0 } & {d^\prime_0 } & 0  \\
   {\sigma e^{i\beta } b^\prime_0 } & 0 & { - d^\prime_0 }  \\
\end{array}} \right),
\label{Eq:C2 Inverted}
\end{equation}
%%%%%%%%%%%%%%%%%%%%
where $\sigma e_0/m_0$  = $1 -2e^{2i\alpha}\eta ^2/(2 - p\eta)+ {\mathcal O}(\varepsilon^2)$ from $s=1$
to give $m_3=0$ from Eq.(\ref{Eq:e from mu-tau symmetric det(Mnu) 2}).

\begin{figure}[!htbp]
\begin{flushleft}
%\includegraphics*[20mm,80mm][200mm,265mm]{phase.eps}
%\includegraphics*[20mm,90mm][200mm,265mm]{phase.eps}
%80==>90 title moves closer to figure
%\includegraphics*[20mm,90mm][200mm,265mm]{phase.eps}
%\includegraphics*[20mm,90mm][200mm,235mm]{phase.eps}
%265==>235 figure moves upward
\includegraphics*[30mm,193mm][195mm,255.2mm]{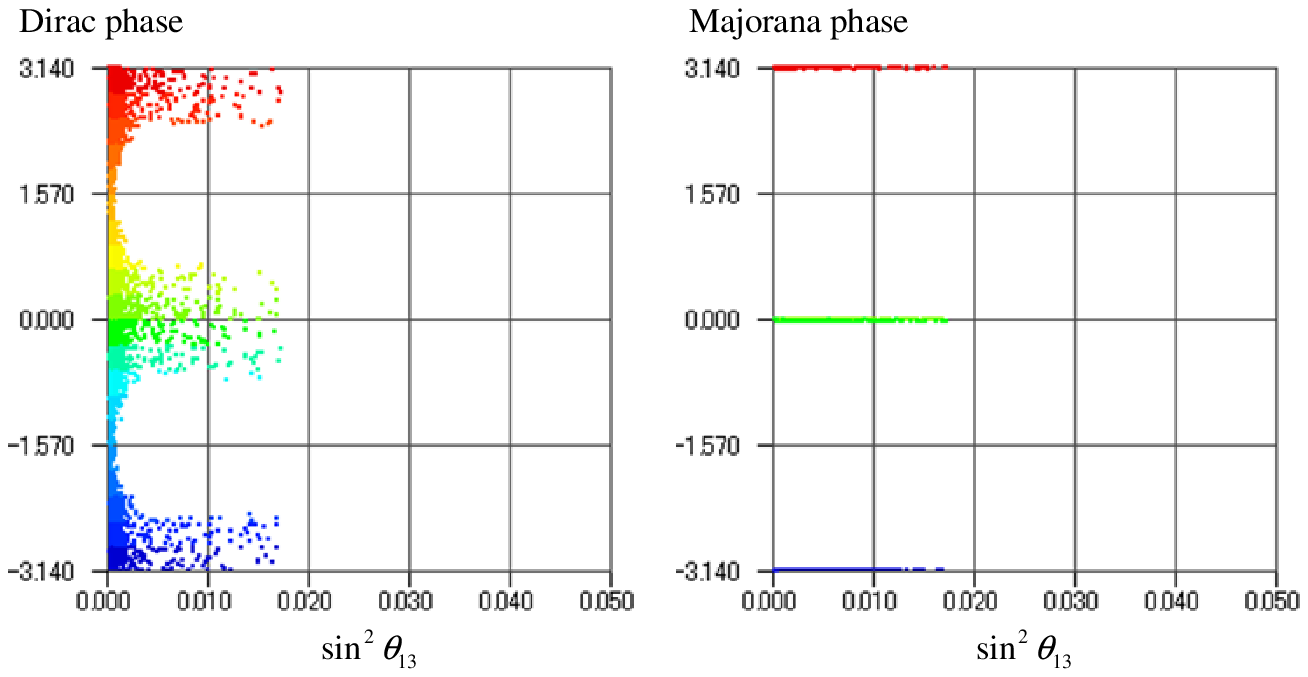}
\end{flushleft}
\vspace{-4mm}
\caption{The same as in FIG.\ref{Fig:C1-N} but 
for the inverted mass hierarchy with $m_1\approx m_2$ given by $M^{(C2)IA}_\nu$.}
\label{Fig:C2-IA}
\end{figure}

We obtain that
%%%%%%%%%%%%%%%%%%%%
\begin{eqnarray}
\tan 2\theta _{12} e^{i\rho }  &\approx& 
\frac{\left[ {4b^\prime_0 \cos \beta  + 
\eta\left(
d^\prime_0 e^{ - i\alpha } 
+
b^\prime_0pe^{ i\beta }
\right)
} \right]\frac{\varepsilon }{\eta } + 2m_0\left( {\Delta  + i\gamma } \right)e^{i\alpha } }{\sqrt 2 m_0p},
\nonumber\\
\tan 2\theta _{13} e^{ - i\delta }  &\approx&  \sqrt 2 \sigma \eta e^{i\alpha }.
\label{Eq:C2 inverted-theta12-theta13}
\end{eqnarray}
%%%%%%%%%%%%%%%%%%%%
Owing to the phenomenological constraint of 
$\sin^22\theta_{12}={\mathcal O}(1)$, $ \eta  \approx  \varepsilon $ and $b^\prime_0\neq 0$ as well as 
$\vert\cos\beta\vert={\mathcal O}(1)$ should be satisfied.
The phases becomes
%%%%%%%%%%%%%%%%%%%%
\begin{equation}
\tan\rho \approx 
\frac{\left( {\varepsilon d^\prime_0  - 2m_0\Delta } \right)\sin \alpha  - 2m_0\gamma \cos \alpha +\varepsilon b^\prime_0 p\sin\beta}
{4b^\prime_0 \frac{\varepsilon }{\eta }\cos \beta  + {\left( {\varepsilon d^\prime_0  + 2m_0\Delta } \right)\cos \alpha  - 2m_0\gamma \sin \alpha- \varepsilon b^\prime_0 p\cos\beta}},
\quad
\delta  \approx  - \alpha.
\label{Eq:C2 inverted-delta-rho}
\end{equation}
%%%%%%%%%%%%%%%%%%%%
Because of $ \eta  \approx  \varepsilon $ and $\vert\cos\beta\vert={\mathcal O}(1)$, 
Eq.(\ref{Eq:C2 inverted-delta-rho}) gives $\rho\approx 0$.
Since masses are calculated to be:
%%%%%%%%%%%%%%%%%%%%
\begin{eqnarray}
 m_1 e^{ - 2i\phi _1 }  &\approx& 
e^{i\rho } \left( {\frac{{4\cos \rho  - p\eta e^{i\rho} }}{2} - \frac{{\sqrt 2 \varepsilon b^\prime_0 e^{i\beta } }}{{\sin 2\theta _{12} }}} \right)m_0,
\nonumber\\
 m_2 e^{ - 2i\phi _3 }  &\approx& 
e^{i\rho } \left( {\frac{{4\cos \rho  - p\eta e^{i\rho}}}{2} + \frac{{\sqrt 2 \varepsilon b^\prime_0 e^{i\beta }}}{{\sin 2\theta _{12} }}} \right)m_0.
\label{C2 Inverted-phi2-phi3}
\end{eqnarray}
%%%%%%%%%%%%%%%%%%%%
Majorana phases become
%%%%%%%%%%%%%%%%%%%%
\begin{equation}
\phi _1  \approx  \phi _2  \approx  - \frac{\rho}{2}.
\label{Eq:C2 inverted-MajoranaPhase}
\end{equation}
%%%%%%%%%%%%%%%%%%%%
Therefore, Majorana CP violation is characterized by
%%%%%%%%%%%%%%%%%%%%
\begin{equation}
\phi_{CP} \approx 0,
\label{Eq:C2 inverted-CPMajoranaPhase}
\end{equation}
%%%%%%%%%%%%%%%%%%%%
which is shown in FIG.\ref{Fig:C2-IA}.

%%%%%%%%%%%%%%%%%%%%%%%%%%%%%%%%%%%%%%%%%%%%%%%%%%%%%%%%%%%%%%%%
\subsubsection{\label{subsubsec:C2)I-B}Inverted mass hierarchy I B ($m_1\approx m_2$)}
%%%%%%%%%%%%%%%%%%%%%%%%%%%%%%%%%%%%%%%%%%%%%%%%%%%%%%%%%%%%%%%%
Our mass matrix $M_\nu$ can be parameterized by
%%%%%%%%%%%%%%%%%%%%
\begin{equation}
M^{(C2)IB}_\nu = m_0\left( {\begin{array}{*{20}c}
   {-2 + p\eta } & {e^{i\alpha } \eta } & { - \sigma e^{i\alpha } \eta }  \\
   {e^{i\alpha } \eta } & 1 & {e_0}/{m_0}  \\
   { - \sigma e^{i\alpha } \eta } & {e_0}/{m_0} & 1  \\
\end{array}} \right) + \varepsilon\left( {\begin{array}{*{20}c}
   0 & {e^{i\beta } b^\prime_0 } & {\sigma e^{i\beta } b^\prime_0 }  \\
   {e^{i\beta } b^\prime_0 } & {d^\prime_0 } & 0  \\
   {\sigma e^{i\beta } b^\prime_0 } & 0 & { - d^\prime_0 }  \\
\end{array}} \right),
\label{Eq:C2 Inverted-2}
\end{equation}
%%%%%%%%%%%%%%%%%%%%
where $\sigma e_0/m_0$  = $1 +2e^{2i\alpha}\eta ^2/(2 - p\eta)+ {\mathcal O}(\varepsilon^2)$ from $s=-1$
to give $m_3=0$ from Eq.(\ref{Eq:e from mu-tau symmetric det(Mnu) 2}).
Similarly to the relation between 
Eq.(\ref{Eq:C1 Inverted1}) and  Eq.(\ref{Eq:C1 Inverted1-2}), 
the sign of $a_0$ differs from the one for Eq.(\ref{Eq:C2 Inverted}).

\begin{figure}[!htbp]
\begin{flushleft}
%\includegraphics*[20mm,80mm][200mm,265mm]{phase.eps}
%\includegraphics*[20mm,90mm][200mm,265mm]{phase.eps}
%80==>90 title moves closer to figure
%\includegraphics*[20mm,90mm][200mm,265mm]{phase.eps}
%\includegraphics*[20mm,90mm][200mm,235mm]{phase.eps}
%265==>235 figure moves upward
\includegraphics*[30mm,193mm][195mm,255.2mm]{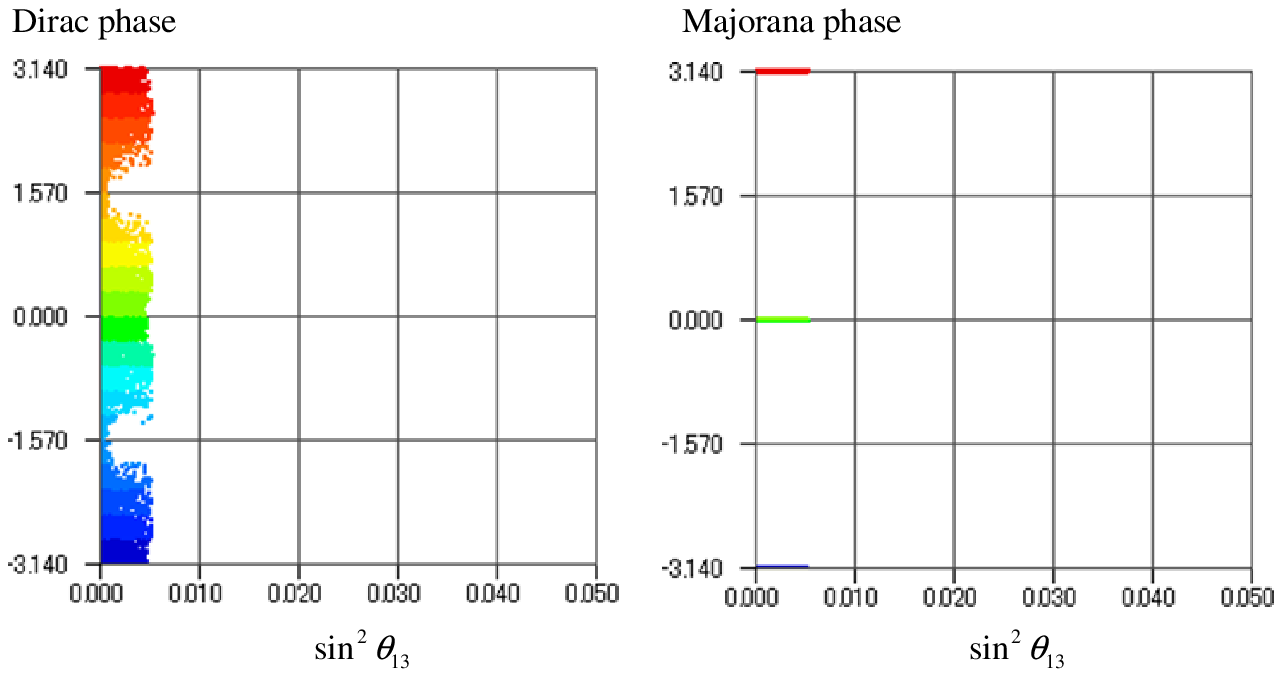}
\end{flushleft}
\vspace{-4mm}
\caption{The same as in FIG.\ref{Fig:C1-N} but  
for the inverted mass hierarchy with $m_1\approx m_2$ given by $M^{(C2)IB}_\nu$.}
\label{Fig:C2-IB}
\end{figure}

We obtain that
\begin{eqnarray}
\tan 2\theta _{12} e^{i\rho }  &\approx& 
-\frac{{\left[ 4b^\prime_0 i\sin \beta  - 
\eta \left(d^\prime_0 e^{ - i\alpha }-b^\prime_0 pe^{ i\beta }\right)
\right]\frac{\varepsilon }{\eta } + 2m_0\left( {\Delta  + i\gamma } \right)e^{i\alpha } }}
{{\sqrt 2 m_0p}},
\nonumber\\
\tan 2\theta _{13} e^{ - i\delta }  &\approx& 
 -\sqrt 2 \sigma \eta e^{i\alpha }.
\label{Eq:C2 inverted-2-theta12-theta13}
\end{eqnarray}
%%%%%%%%%%%%%%%%%%%%
Owing to the phenomenological constraint of 
$\sin^22\theta_{12}={\mathcal O}(1)$, $ \eta  \approx  \varepsilon $  and $b^\prime_0\neq 0$ as well as 
$\vert\sin\beta\vert={\mathcal O}(1)$ should be satisfied.
The phases becomes
%%%%%%%%%%%%%%%%%%%%
\begin{equation}
\tan\rho \approx 
- \frac{4\frac{\varepsilon}{\eta}b^\prime_0 \sin \beta  + 2m_0\gamma \cos \alpha  + \left( {\varepsilon d^\prime_0  + 2m_0\Delta } \right) \sin \alpha-\varepsilon b^\prime_0p\sin\beta }
 {\left( {\varepsilon d^\prime_0  - 2m_0\Delta } \right) \cos \alpha  + 2m_0\gamma  \sin \alpha+\varepsilon b^\prime_0p\cos\beta},
\quad
\delta  \approx  - \alpha.
\label{Eq:C2 inverted-2-delta-rho}
\end{equation}
%%%%%%%%%%%%%%%%%%%%
Similarly to Eq.(\ref{Eq:C2 inverted-delta-rho}), Eq.(\ref{Eq:C2 inverted-2-delta-rho}) gives 
$\rho\approx \pm\pi/2$.  Since masses are calculated to be:
%%%%%%%%%%%%%%%%%%%%
\begin{eqnarray}
 m_1 e^{ - 2i\phi _1 }  &\approx&
e^{i\rho } \left( {\frac{{ - 4i\sin \rho  + p\eta e^{i\rho } }}{2}m_0  - 
\frac{\sqrt 2 \varepsilon b'_0 e^{i\beta }}{{\sin 2\theta _{12} }}} \right),
\nonumber\\
 m_2 e^{ - 2i\phi _3 }  &\approx &
e^{i\rho } \left( {\frac{{ - 4i\sin \rho  + p\eta e^{i\rho } }}{2}m_0  + 
\frac{\sqrt 2 \varepsilon b'_0 e^{i\beta }}{{\sin 2\theta _{12} }}} \right).
\label{C1 Inverted-2-phi2-phi3}
\end{eqnarray}
%%%%%%%%%%%%%%%%%%%%
Majorana phases become
%%%%%%%%%%%%%%%%%%%%
\begin{equation}
\phi _1  \approx \phi _2  \approx  
 - \frac{1}{2}\left(\rho-\frac{\pi }{2} \right).
\label{Eq:C2 inverted-2-MajoranaPhase}
\end{equation}
%%%%%%%%%%%%%%%%%%%%
Therefore, Majorana CP violation is characterized by
%%%%%%%%%%%%%%%%%%%%
\begin{equation}
\phi_{CP} \approx 0,
\label{Eq:C2 inverted-2-CPMajoranaPhase}
\end{equation}
%%%%%%%%%%%%%%%%%%%%
as shown in FIG.\ref{Fig:C2-IB}.

%%--------------------------------
%%
%%--------------------------------
%%%%%%%%%%%%%%%%%%%%%%%%%%%%%%%%%%%%%%%%%%%%%%%%%%%%%%%%%%%%%%%%%%%%%%%%%%%%%%%%
\section{\label{sec:6}Summary and Discussions}
To discuss leptonic CP violation as direct effects from phases of flavor neutrino masses
in a model-independent way, we have focused the general parameterization of $U_{PMNS}$ 
that can take care of redundant phases 
originally arising from in the arbitrariness in phases of flavor neutrino masses.  
As a result, we have found that the Dirac CP phase $\delta_{CP}$ is determined from
%%%%%%%%%%%%%%%%%%%%
\begin{equation}
\delta_{CP} = \rho+\delta~{\rm with}~\rho={\rm arg}(X)~{\rm and}~\delta=-{\rm arg}(Y),
\label{Eq:delta-rho delta from X-Y}
\end{equation}
%%%%%%%%%%%%%%%%%%%%
where $X$ and $Y$ are described by the flavor neutrino masses:
%%%%%%%%%%%%%%%%%%%%
\begin{eqnarray}
&&c_{13}X = \left( {{c_{23}} + \sigma {s_{23}}} \right)\left( {{B_ + }\cos \gamma  + i{B_ - }\sin \gamma } \right)
\nonumber\\
&&\qquad + \left( {{c_{23}} - \sigma {s_{23}}} \right)\left( {{B_ - }\cos \gamma  + i{B_ + }\sin \gamma } \right),
\nonumber\\
&& Y = \sigma \left[ \begin{array}{l}
 \left( {{c_{23}} + \sigma {s_{23}}} \right)\left( {{B_ - }\cos \gamma  + i{B_ + }\sin \gamma } \right) \\ 
  - \left( {{c_{23}} - \sigma {s_{23}}} \right)\left( {{B_ + }\cos \gamma  + i{B_ - }\sin \gamma } \right) \\ 
 \end{array} \right]
\label{Eq:X and Y}
\end{eqnarray}
%%%%%%%%%%%%%%%%%%%%
with
%%%%%%%%%%%%%%%%%%%%
\begin{equation}
 B_ +   + B_ - = \sum\limits_{f = e}^\tau  {M_{ef}^ *  M_{f\mu } },
\qquad
 B_ +   - B_ - = -\sigma\sum\limits_{f = e}^\tau  {M_{ef}^ *  M_{f\tau } },
\label{Eq:Bplus and Bminus}
\end{equation}
%%%%%%%%%%%%%%%%%%%%
as can be derived from Eq.(\ref{Eq:App X Y from M^daggerM}).  Under the approximate $\mu$-$\tau$ symmetry, 
we have obtained $X\approx \sqrt{2}B_+$ for the category (C1) and 
$Y\approx -\sqrt{2}\sigma B_+$ for the category (C2) and other quantities 
shows complicated relations among terms of order $\varepsilon$.  More precisely, 
the model is characterized by the two phases $\alpha$ and $\beta$ introduced as phases of $M^{(\pm)}_{e\mu}$ and 
simple relations between the CP phases $\rho$ and $\delta$ and 
our specific phases $\alpha$ and $\beta$ turn out to arise, in the category (C1),
%%%%%%%%%%%%%%%%%%%%
\begin{enumerate}
\item %'±'±'É•¶'ð'¢'ê'é
for the normal mass hierarchy, $\rho\approx\alpha$ and $\delta\approx\beta$,
\item 
for the inverted  mass hierarchy with $m_1\approx m_2$, $\rho\approx 0$ or $\pm \pi/2$ and $\delta\approx-\beta$,
\end{enumerate}
%%%%%%%%%%%%%%%
and, in the category (C2),
%%%%%%%%%%%%%%%%%%%%
\begin{enumerate}
\item 
for the inverted  mass hierarchy with $m_1\approx m_2$, $\delta\approx-\alpha$.
\end{enumerate}
%%%%%%%%%%%%%%%
Other cases do not show such simple relations.  

Majorana CP violation can only be enhanced in the normal mass hierarchy and 
the inverted mass hierarchy with $m_1\approx -m_2$. 
There are two kinds of the 
inverted mass hierarchy depending on the relative sign of $m_1$ and $m_2$, 
namely, with $m_1\approx m_2$ and $m_1\approx -m_2$. Majorana CP violation
is suppressed for the case 
with $m_1\approx m_2$ and is much enhanced 
for the case with $m_1\approx -m_2$.  Numerically, the enhanced size of the CP violating Majorana phase
is given by $-\pi/4\mapleq\phi_{CP}\mapleq \pi/4$ (mod $\pi$).  
Maximal Dirac CP violation can arise for
%%%%%%%%%%%%%%%%%%%%
\begin{enumerate}
\item 
$\sin^2\theta_{13} \mapleq 0.001$ in the inverted mass hierarchies I A and I B realized for the category (C2),
\item 
$\sin^2\theta_{13} \mapleq 0.01$ in the inverted mass hierarchies II A and II B realized for the category (C1),
\item 
$\sin^2\theta_{13} \mapleq 0.04$ in the normal mass hierarchy and the inverted mass hierarchy I A both realized for the category (C1).
\end{enumerate}
%%%%%%%%%%%%%%%
On the other hand, maximal Majorana CP violation is only possible to arise for $\sin^2\theta_{13} \mapleq 0.03$
in the normal mass hierarchy realized for the category (C1).  
It has been noted that these predictions do not depend on the choice of our specific phases because of the 
rephasing invariance in our formalism. 

Finally, if $\det(M_\nu)=0$ is the result of the minimal seesaw model, 
these predictions are valid at the 
seesaw scale and are modified at the weak scale by the renormalization \cite{RGE}, whose effects will be 
evaluated in the forthcoming paper \cite{Future-work}.

%%%%% Acknowledgements %%%%%%%%%%%%%%%%%%%
\section*{Acknowledgements}
The authors would like to thank T. Kitabayashi for valuable comments and discussions.
%%%%% Appendix %%%%%%%%%%%%%%%%%%%
\appendix
\section{\label{sec:Appendix-rephasing} Rephasing invariance in Dirac and Majorana phases}

Let us demonstrate the rephasing invariance of $\delta_{CP}$ and $\phi_{CP}$, which are not trivial 
when these are expressed in flavor neutrino masses, 
by considering the induced changes 
in Eqs.(\ref{Eq:App rho delta from X-Y}) for $\rho$ and $\delta$ and 
(\ref{Eq:App Mass from M}) for Majorana phases.  One particularly chooses 
some of phases of flavor neutrino masses to be real by removing theirs phases by the rephasing.
To see the rephasing invariance, we first show how $\delta$ and $\rho$ vary 
with the rephasing.
Rephasing the charged leptons ($\ell$) is caused by $\ell^\prime =U(\theta)\ell$, 
where $U(\theta)={\rm diag.}(e^{i\theta_e},e^{i\theta_\mu},e^{i\theta_\tau})$,
which in turn calls for the redefinition of the flavor neutrinos: 
$\nu^\prime_f = U(\theta) \nu_f$.  As a result,
the mass term $\nu^T_f M_\nu\nu_f$ is equivalent to $\nu^{\prime T}_f M^\prime_\nu\nu^\prime_f$
 with $M^\prime_\nu$ defined by
%%%%%%%%%%%%%%%%%%%%
\begin{eqnarray}
M^\prime_\nu   
= \left( {\begin{array}{*{20}c}
   {e^{ - 2i\theta _e } M_{ee} } & {e^{ - i\left( {\theta _e  + \theta _\mu  } \right)} M_
{e\mu } } & {e^{ - i\left( {\theta _e  + \theta _\tau  } \right)} M_{e\tau } }  \\
   {e^{ - i\left( {\theta _e  + \theta _\mu  } \right)} M_{e\mu } } & {e^{ - 2i\theta _\mu  }
 M_{\mu \mu } } & {e^{ - i\left( {\theta _\mu   + \theta _\tau  } \right)} M_{\mu \tau } }
  \\
   {e^{ - i\left( {\theta _e  + \theta _\tau  } \right)} M_{e\tau } } & {e^{ - i\left( {
\theta _\mu   + \theta _\tau  } \right)} M_{\mu \tau } } & {e^{ - 2i\theta _\tau  } M_{\tau 
\tau } }  \\
\end{array}} \right).
\label{Eq:App M_nu-dash}
\end{eqnarray}
%%%%%%%%%%%%%%%%%%%%
This mass matrix $M^\prime_\nu$ yields
%%%%%%%%%%%%%%%%%%%%
\begin{eqnarray}
X^\prime  &=& e^{i\left( {\theta _e  - \frac{{\theta _\mu   + \theta _\tau  }}{2}} \right)} X,
\qquad
Y^\prime  = e^{i\left( {\theta _e  - \frac{{\theta _\mu   + \theta _\tau  }}{2}} \right)} Y,
\label{Eq:App X-Y-dash}
\end{eqnarray}
%%%%%%%%%%%%%%%%%%%%
where
%%%%%%%%%%%%%%%%%%%%
\begin{eqnarray}
&&\gamma^\prime = \gamma  - \frac{{\theta _\tau   - \theta _\mu  }}{2},
\label{Eq:App gamma-dash}
\end{eqnarray}
%%%%%%%%%%%%%%%%%%%%
derived from Eq.(\ref{Eq:App MixingAngle23 from M^daggerM}) is used in $X^\prime$ and $Y^\prime$, from which
%%%%%%%%%%%%%%%%%%%%
\begin{eqnarray}
 \delta^\prime  &=& \delta -\left( \theta _e  - \frac{\theta _\mu   + \theta _\tau  }{2}\right),
 \qquad
 \rho^\prime  = \rho +\theta _e  - \frac{\theta _\mu   + \theta _\tau  }{2},
\label{Eq:App delta-rho-dash}
\end{eqnarray}
%%%%%%%%%%%%%%%%%%%%
are obtained.  Next, it is, thus, confirmed that $\delta_{CP}$ defined by 
%%%%%%%%%%%%%%%%%%%%
\begin{equation}
\delta_{CP}=\delta+\rho,
\label{Eq:App delta-CP}
\end{equation}
%%%%%%%%%%%%%%%%%%%%
is a rephasing-invariant quantity.  Similarly, we find that Majorana phases become
%%%%%%%%%%%%%%%%%%%%
\begin{equation}
\varphi^\prime_i  = \varphi _i  + \frac{{\theta _\mu   + \theta _\tau  }}{2},
\label{Eq:App majorana-dash}
\end{equation}
%%%%%%%%%%%%%%%%%%%%
for $i=1,2,3$, by using
%%%%%%%%%%%%%%%%%%%%
\begin{eqnarray}
&& \lambda^\prime_1  = e^{ - i\left( {\theta _\mu   + \theta _\tau  } \right)} \lambda _1,
 \qquad
 \lambda^\prime_2  = e^{ - i\left( {\theta _\mu   + \theta _\tau  } \right)} \lambda _2 ,
 \qquad
 \lambda^\prime_3  = e^{ - i\left( {\theta _\mu   + \theta _\tau  } \right)} \lambda _3 ,
\nonumber\\
&& x^\prime = e^{ - i\left( {\theta _\mu   + \theta _\tau  } \right)} x.
\label{Eq:App lamda-x-y-dash}
\end{eqnarray}
%%%%%%%%%%%%%%%%%%%%
The physical Majorana phases defined by $\varphi^\prime_i-\varphi^\prime_j$ ($i\neq j$) for 
$i,j=1,2,3$  turn out to be rephasing-invariant.

It is instructive to note that there are three typical forms of 
the PMNS unitary matrix depending on how the flavor neutrinos are redefined:
\begin{enumerate}
\item $U_{PMNS}$ with $\delta$, $\rho$ and $\gamma$
%%%%%%%%%%%%%%%%%%%%
\begin{eqnarray}
&&
\left( {\begin{array}{*{20}c}
   {c_{12} c_{13} } & {s_{12} c_{13} e^{i\rho } } & {s_{13} e^{ - i\delta } }  \\
   { - \left( \begin{array}{l}
 c_{23} s_{12} e^{ - i\rho }  \\ 
 + s_{23} c_{12} s_{13} e^{i\delta }  \\ 
 \end{array} \right)e^{i\gamma } } & {\left( \begin{array}{l}
 c_{23} c_{12}  \\ 
 - s_{23} s_{12} s_{13}  \\ 
 \end{array} \right)e^{i\gamma } } & {s_{23} c_{13} e^{i\gamma } }  \\
   {\left( \begin{array}{l}
 s_{23} s_{12} e^{ - i\rho }  \\ 
  - c_{23} c_{12} s_{13} e^{i\delta }  \\ 
 \end{array} \right)e^{ - i\gamma } } & { - \left( \begin{array}{l}
 s_{23} c_{12} \\ 
  + c_{23} s_{12} s_{13} e^{i\left( {\delta  + \rho } \right)}  \\ 
 \end{array} \right)e^{ - i\gamma } } & {c_{23} c_{13} e^{ - i\gamma } }  \\
\end{array}} \right)
\nonumber\\
&&
\cdot\left( {\begin{array}{*{20}c}
   {e^{i\varphi _1 } } & 0 & 0  \\
   0 & {e^{i\varphi _2 } } & 0  \\
   0 & 0 & {e^{i\varphi _3 } }  \\
\end{array}} \right),
\label{Eq:PMNSMatrices1}
\end{eqnarray}
%%%%%%%%%%%%%%%%%%%%
for
%%%%%%%%%%%%%%%%%%%%
\begin{eqnarray}
&&
M_\nu = \left( {\begin{array}{*{20}c}
   M_{ee} & M_{e\mu} & M_{e\tau}  \\
   M_{e\mu} & M_{\mu\mu} & M_{\mu\tau}  \\
   M_{e\tau} & M_{\mu\tau} & M_{\tau\tau}  \\
\end{array}} \right),
\end{eqnarray}
%%%%%%%%%%%%%%%%%%%%
\item $U_{PMNS}$ with $\delta$ and $\rho$
%%%%%%%%%%%%%%%%%%%%
\begin{eqnarray}
&&
 \left( {\begin{array}{*{20}c}
   {c_{12} c_{13} }  \\
   { - c_{23} s_{12} e^{ - i\rho }  - s_{23} c_{12} s_{13} e^{i\delta } }  \\
   {s_{23} s_{12} e^{ - i\rho }  - c_{23} c_{12} s_{13} e^{i\delta } }  \\
\end{array}{\rm{ }}\begin{array}{*{20}c}
   {s_{12} c_{13} e^{i\rho } }  \\
   {c_{23} c_{12}  - s_{23} s_{12} s_{13} e^{i\left( {\delta  + \rho } \right)} }  \\
   { - s_{23} c_{12}  - c_{23} s_{12} s_{13} e^{i\left( {\delta  + \rho } \right)} }  \\
\end{array}{\rm{ }}\begin{array}{*{20}c}
   {s_{13} e^{ - i\delta } }  \\
   {s_{23} c_{13} }  \\
   {c_{23} c_{13} }  \\
\end{array}} \right)
\nonumber\\
&&\cdot
\left( {\begin{array}{*{20}c}
   {e^{i\varphi _1 } } & 0 & 0  \\
   0 & {e^{i\varphi _2 } } & 0  \\
   0 & 0 & {e^{i\varphi _3 } }  \\
\end{array}} \right),
\label{Eq:PMNSMatrices2}
\end{eqnarray}
%%%%%%%%%%%%%%%%%%%%
for
%%%%%%%%%%%%%%%%%%%%
\begin{eqnarray}
&&
M^{Intermediate}_\nu = \left( {\begin{array}{*{20}c}
   M_{ee} & {e^{i {\gamma }} M_{e\mu}} & {e^{-i \gamma } M_{e\tau}}  \\
   {e^{i\gamma } M_{e\mu}} & {e^{2i\gamma } M_{\mu\mu}} & M_{\mu\tau}  \\
   {e^{i\gamma } M_{e\tau}} & M_{\mu\tau} & {e^{ - 2i\gamma } M_{\tau\tau}}  \\
\end{array}} \right),
\end{eqnarray}
%%%%%%%%%%%%%%%%%%%%
\item $U_{PMNS}$ with $\delta_{CP}=\delta+\rho$, $\phi_1=\varphi_1-\rho$ and $\phi_{2,3}=\varphi_{2,3}$ 
%%%%%%%%%%%%%%%%%%%%
\begin{eqnarray}
&&
 \left( {\begin{array}{*{20}c}
   {c_{12} c_{13} }  \\
   { - c_{23} s_{12}  - s_{23} c_{12} s_{13} e^{i \delta_{CP}} }  \\
   {s_{23} s_{12}  - c_{23} c_{12} s_{13} e^{i\delta_{CP}} }  \\
\end{array}{\rm{ }}\begin{array}{*{20}c}
   {s_{12} c_{13} }  \\
   {c_{23} c_{12}  - s_{23} s_{12} s_{13} e^{i\delta_{CP}} }  \\
   { - s_{23} c_{12}  - c_{23} s_{12} s_{13} e^{i\delta_{CP}} }  \\
\end{array}{\rm{ }}\begin{array}{*{20}c}
   {s_{13} e^{ - i\delta_{CP}} }  \\
   {s_{23} c_{13} }  \\
   {c_{23} c_{13} }  \\
\end{array}} \right)
\nonumber\\
&&
\cdot
\left( {\begin{array}{*{20}c}
   {e^{i\phi _1} } & 0 & 0  \\
   0 & {e^{i\phi _2 } } & 0  \\
   0 & 0 & {e^{i\phi _3 } }  \\
\end{array}} \right),
\label{Eq:PMNSMatrices3}
\end{eqnarray}
%%%%%%%%%%%%%%%%%%%%
for
%%%%%%%%%%%%%%%%%%%%
\begin{eqnarray}
&&
M^{PDG}_\nu = \left( {\begin{array}{*{20}c}
   {e^{2i\rho }M_{ee}} & {e^{i\left( {\rho  + \gamma } \right)} M_{e\mu}} & {e^{i\left( {\rho  - \gamma } \right)} M_{\tau\tau}}  \\
   {e^{i\left( {\rho  + \gamma } \right)} M_{e\mu}} & {e^{2i\gamma } M_{\mu\mu}} & M_{\mu\tau}  \\
   {e^{i\left( {\rho  - \gamma } \right)} M_{e\tau}} & M_{\mu\tau} & {e^{ - 2i\gamma } M_{\tau\tau}}  \\
\end{array}} \right).
\label{Eq:M_nu PDG}
\end{eqnarray}
%%%%%%%%%%%%%%%%%%%%
\end{enumerate}
%%%%%%%%%%%%%%%%%%%%

%%%%%%%%%%%%%%%%%%%%%%%%%%%%%%%%%%%%%%%%%%%%%%%%%%%%%%%%%%%%%%%%%%%%%%%%%%%%%%%%%%%%%%%%%%%%%%%%
\section{\label{sec:Appendix-massless} Massless Neutrino in Models with $\det$(M$_\nu$)=0}

In this appendix, we discuss how one massless neutrino arises from our mass formula when det($M_\nu$)=0 is 
applied to it.  Because corrections offered by $M^{(-)}_\nu$ turn out to be ${\mathcal{O}}(\varepsilon^2)$, 
we may evaluate det($M_\nu$)=0 in the $\mu$-$\tau$ symmetric limit to get valid results up to ${\mathcal(O)}(\varepsilon)$.
The relations determined by det($M_\nu$)=0 are either Eq.(\ref{Eq:e from mu-tau symmetric det(Mnu) 1}) or
Eq.(\ref{Eq:e from mu-tau symmetric det(Mnu) 2}).

For the category (C1) with $\sin\theta_{13}=0$, neutrino masses are given by
%%%%%%%%%%%%%%%%%%%%
\begin{eqnarray}
m_1 e^{ - 2i\varphi _1 }  &=& \frac{{e^{2i\rho } a  + d  - \sigma e }}{2} - \frac{{\sqrt 2 e^{i\rho } b }}{{\sin 2\theta _{12} }},
\nonumber\\
m_2 e^{ - 2i\varphi _2 }  &=& \frac{{e^{2i\rho } a  + d  - \sigma e }}{2} + \frac{{\sqrt 2 e^{i\rho } b }}{{\sin 2\theta _{12} }},
\nonumber\\
m_3 e^{ - 2i\varphi _3 }  &=& d  + \sigma e. 
\label{Eq:App masses for C1 sin13=0}
\end{eqnarray}
%%%%%%%%%%%%%%%%%%%%
If Eq.(\ref{Eq:e from mu-tau symmetric det(Mnu) 1}) is satisfied, $m_3=0$ is derived.
On the other hand, if Eq.(\ref{Eq:e from mu-tau symmetric det(Mnu) 2}) is satisfied, 
we further evaluate $m_{1,2}$.  By evaluating $\sin 2\theta_{12}$ from
$\tan^22\theta_{12}$ in Eq.(\ref{Eq:App MixingAngle from M}), where
Eq.(\ref{Eq:e from mu-tau symmetric det(Mnu) 2}) is used to replaces $b^2$ in $\tan^22\theta_{12}$, 
we reach the relation
%%%%%%%%%%%%%%%%%%%%
\begin{equation}
\frac{\sqrt 2 e^{i\rho } b }{\sin 2\theta _{12}} = \frac{\sqrt {z^2 }}{2},
\label{Eq:App x div sin(2theta12) C1}
\end{equation}
%%%%%%%%%%%%%%%%%%%%
where
%%%%%%%%%%%%%%%%%%%%
\begin{equation}
z = e^{2i\rho } a + d  - \sigma e.
\label{Eq:App sqrt(z)=pm z C1}
\end{equation}
%%%%%%%%%%%%%%%%%%%%
We then find that
%%%%%%%%%%%%%%%%%%%%
\begin{eqnarray}
\left. \begin{array}{l}
 m_1 e^{ - 2i\varphi _1 }  \\ 
 m_2 e^{ - 2i\varphi _2 }  \\ 
 \end{array} \right\} 
 &=& \left\{ \begin{array}{l}
 0 \\ 
 e^{2i\rho } a  + d  - \sigma e  \\ 
 \end{array} \right.~ \left( {\sqrt {z^2 }  = z} \right),
 \nonumber\\
 &=&\left\{ \begin{array}{l}
 e^{2i\rho } a  + d  - \sigma e  \\ 
 0 \\ 
 \end{array} \right.~\left( {\sqrt {z^2 }  =  - z} \right),
\nonumber\\
m_3 e^{ - 2i\varphi _3 }  &=& d  + \sigma e.
\label{Eq:App m123 for C1 sin13=0}
\end{eqnarray}
%%%%%%%%%%%%%%%%%%%%
Since the case with $m_2=0$ is phenomenologically excluded, the condition of 
$\sqrt {z^2 }  = z$ should be satisfied.

For the category (C2) with $\sin\theta_{12}=0$, neutrino masses are given by
%%%%%%%%%%%%%%%%%%%%
\begin{eqnarray}
 m_1 e^{ - 2i\varphi _1 }  &=& \frac{{e^{2i\rho } }}{2}\left( {a  + e^{2i\delta } \left( {d  - \sigma e } \right) + \frac{{a - e^{2i\delta } \left( {d  - \sigma e } \right)}}{{\cos 2\theta _{13} }}} \right),
\nonumber\\
 m_2 e^{ - 2i\varphi _2 }  &=& d  + \sigma e, 
\nonumber\\
 m_3 e^{ - 2i\varphi _3 }  &=&  \frac{{e^{2i\rho } }}{2}\left( {a  + e^{2i\delta } \left( {d  - \sigma e } \right) - \frac{{a - e^{2i\delta } \left( {d  - \sigma e } \right)}}{{\cos 2\theta _{13} }}} \right).
\label{Eq:App masses for C2 sin12=0}
\end{eqnarray}
%%%%%%%%%%%%%%%%%%%%
Similarly for the category (C1), 
if Eq.(\ref{Eq:e from mu-tau symmetric det(Mnu) 1}) is satisfied, $m_2=0$ is derived.
On the other hand, if Eq.(\ref{Eq:e from mu-tau symmetric det(Mnu) 2}) is satisfied, 
we further evaluate $m_{1,2}$.  
In the manner used to derive Eq.(\ref{Eq:App x div sin(2theta12) C1}) in the category (C1), 
we reach the relation
%%%%%%%%%%%%%%%%%%%%
\begin{equation}
\frac{1}{{\cos 2\theta _{13} }} = k\frac{\sqrt{z^2}}{{\left( {d  - \sigma e } \right)e^{i\delta }  - a e^{ - i\delta }}},
\label{Eq:App x div sin(2theta13) C2}
\end{equation}
%%%%%%%%%%%%%%%%%%%%
where $k=\pm 1$ and
%%%%%%%%%%%%%%%%%%%%
\begin{equation}
z = \left( {d  - \sigma e } \right)e^{i\delta }  + a e^{ - i\delta }.
\label{Eq:App sqrt(z)=pm z C2}
\end{equation}
%%%%%%%%%%%%%%%%%%%%
We then find that
%%%%%%%%%%%%%%%%%%%%
\begin{eqnarray}
\left. \begin{array}{l}
 m_1 e^{ - 2i\varphi _1 }  \\ 
 m_3 e^{ - 2i\varphi _3 }  \\ 
  \end{array} \right\} 
&=&
 \left\{ \begin{array}{l}
 0 \\ 
 e^{ - 2i\delta } \left( {a  + e^{2i\delta } \left( {d  - \sigma e } \right)} \right) \\ 
 \end{array} \right.~ \left( {\sqrt {z^2 }  = kz} \right),
\nonumber\\
&=&\left\{ \begin{array}{l}
 e^{2i\rho } \left( {a  + e^{2i\delta } \left( {d  - \sigma e } \right)} \right) \\ 
 0 \\ 
 \end{array} \right.~\left( {\sqrt {z^2 }  =  - kz} \right),
\nonumber\\
m_2 e^{ - 2i\varphi _2 }  &=& d  + \sigma e.
\label{Eq:App m123 for C2 sin12=0}
\end{eqnarray}
%%%%%%%%%%%%%%%%%%%%

%%%%%%%%%%%%%%%%%%%%%%%%%%%%%%%%%%%%%%%%%
%%%%% Reference %%%%%%%%%%%%%%%%%%%
\bigskip % extra skip inserted

%%%%% Figures %%%%%%%%%%%%%%%%%%%
\newpage
\centerline{\bf Figures}
%%-------------------------------------------------
%% Figures
%%-------------------------------------------------
%%%%%%%%%%%%%%%%%%%%%%%%%%%%%%%%%%%%%%%%%%%%%%%%%%%%%%%%%%%%%%%%%%%%%%%%%%%%%%%%
%%%%%%%%%%%%%%%%%%%%%%%%%%%%%%%%%%%%%%%%%%%%%%%%%%%%%%%%%%%%%%%%%%%%%%%%%%%%%%%%
%%%%%%%%%%%%%%%%%%%%%%%%%%%%%%%%%%%%%%%%%%%%%%%%%%%%%%%%%%%%%%%%%%%%%%%%%%%%%%%%
%%%%%%%%%%%%%%%%%%%%%%%%%%%%%%%%%%%%%%%%%%%%%%%%%%%%%%%%%%%%%%%%%%%%%%%%%%%%%%%%
%%%%%%%%%%%%%%%%%%%%%%%%%%%%%%%%%%%%%%%%%%%%%%%%%%%%%%%%%%%%%%%%%%%%%%%%%%%%%%%%
%%%%%%%%%%%%%%%%%%%%%%%%%%%%%%%%%%%%%%%%%%%%%%%%%%%%%%%%%%%%%%%%%%%%%%%%%%%%%%%%
%%%%%%%%%%%%%%%%%%%%%%%%%%%%%%%%%%%%%%%%%%%%%%%%%%%%%%%%%%%%%%%%%%%%%%%%%%%%%%%%
%%%%%%%%%%%%%%%%%%%%%%%%%%%%%%%%%%%%%%%%%%%%%%%%%%%%%%%%%%%%%%%%%%%%%%%%%%%%%%%%
%%%%%%%%%%%%%%%%%%%%%%%%%%%%%%%%%%%%%%%%%%%%%%%%%%%%%%%%%%%%%%%%%%%%%%%%%%%%%%%%
%%%%%%%%%%%%%%%%%%%%%%%%%%%%%%%%%%%%%%%%%%%%%%%%%%%%%%%%%%%%%%%%%%%%%%%%%%%%%%%%
%%%%%%%%%%%%%%%%%%%%%%%%%%%%%%%%%%%%%%%%%%%%%%%%%%%%%%%%%%%%%%%%%%%%%%%%%%%%%%%%
%%%%%%%%%%%%%%%%%%%%%%%%%%%%%%%%%%%%%%%%%%%%%%%%%%%%%%%%%%%%%%%%%%%%%%%%%%%%%%%%
%%%%%%%%%%%%%%%%%%%%%%%%%%%%%%%%%%%%%%%%%%%%%%%%%%%%%%%%%%%%%%%%%%%%%%%%%%%%%%%%

\end{document}